\documentclass[twocolumn]{aastex631}

\usepackage{amsmath}

\begin{document}

\title{COSMOS-Web: Unraveling the Evolution of Galaxy Size and Related Properties at \boldmath{$2<z<10$}}

\author[0000-0002-8434-880X]{Lilan Yang}
\affiliation{Laboratory for Multiwavelength Astrophysics, School of Physics and Astronomy, Rochester Institute of Technology, 84 Lomb Memorial Drive, Rochester, NY 14623, USA}

\author[0000-0001-9187-3605]{Jeyhan S. Kartaltepe}
\affiliation{Laboratory for Multiwavelength Astrophysics, School of Physics and Astronomy, Rochester Institute of Technology, 84 Lomb Memorial Drive, Rochester, NY 14623, USA}

\author[0000-0002-3560-8599]{Maximilien Franco}
\affiliation{Université Paris-Saclay, Université Paris Cité, CEA, CNRS, AIM, 91191 Gif-sur-Yvette, France}
\affiliation{The University of Texas at Austin, 2515 Speedway Blvd Stop C1400, Austin, TX 78712, USA}

\author[0000-0002-0786-7307]{Xuheng Ding}
\affiliation{School of Physics and Technology, Wuhan University, Wuhan 430072, China}

\author[0009-0000-7385-3539]{Mark J. Achenbach}
\affiliation{Department of Physics and Astronomy, University of Hawaii at Manoa, 2505 Correa Rd, Honolulu, HI 96822, USA}

\author[0000-0002-0569-5222]{Rafael C. Arango-Toro}
\affiliation{Aix Marseille Univ, CNRS, CNES, LAM, Marseille, France  }

\author[0000-0002-0930-6466]{Caitlin M. Casey}
\affiliation{Department of Physics, University of California, Santa Barbara, Santa Barbara, CA 93109, USA}
\affiliation{The University of Texas at Austin, 2515 Speedway Blvd Stop C1400, Austin, TX 78712, USA}
\affiliation{Cosmic Dawn Center (DAWN), Denmark}

\author[0000-0003-4761-2197]{Nicole E. Drakos}
\affiliation{Department of Physics and Astronomy, University of Hawaii, Hilo, 200 W Kawili St, Hilo, HI 96720, USA}

\author[0000-0002-9382-9832]{Andreas L. Faisst}
\affiliation{Caltech/IPAC, MS 314-6, 1200 E. California Blvd. Pasadena, CA 91125, USA}

\author[0000-0001-9885-4589]{Steven Gillman}
\affiliation{Cosmic Dawn Center (DAWN), Denmark}
\affiliation{DTU-Space, Technical University of Denmark, Elektrovej 327, DK-2800 Kgs. Lyngby, Denmark}

\author[0000-0002-0236-919X]{Ghassem Gozaliasl}
\affiliation{Department of Computer Science, Aalto University, P.O. Box 15400, FI-00076 Espoo, Finland}
\affiliation{Department of Physics, University of, P.O. Box 64, FI-00014 Helsinki, Finland}

\author[0000-0002-1416-8483]{Marc Huertas-Company}
\affil{Instituto de Astrof\'isica de Canarias, La Laguna, Tenerife, Spain}
\affil{Universidad de la Laguna, La Laguna, Tenerife, Spain}
\affil{Universit\'e Paris-Cit\'e, LERMA - Observatoire de Paris, PSL, Paris, France}

\author[0000-0002-8412-7951]{Shuowen Jin}\altaffiliation{Marie Curie Fellow}
\affiliation{Cosmic Dawn Center (DAWN), Denmark}
\affiliation{DTU-Space, Technical University of Denmark, Elektrovej 327, 2800 Kgs. Lyngby, Denmark}

\author[0000-0001-9773-7479]{Daizhong Liu}
\affiliation{Purple Mountain Observatory, Chinese Academy of Sciences, 10 Yuanhua Road, Nanjing 210023, China}

\author[0000-0002-4872-2294]{Georgios Magdis}
\affiliation{Cosmic Dawn Center (DAWN), Denmark} 
\affiliation{DTU-Space, Technical University of Denmark, Elektrovej 327, 2800, Kgs. Lyngby, Denmark}
\affiliation{Niels Bohr Institute, University of Copenhagen, Jagtvej 128, DK-2200, Copenhagen, Denmark}

\author[0000-0002-6085-3780]{Richard Massey}
\affiliation{Department of Physics, Centre for Extragalactic Astronomy, Durham University, South Road, Durham DH1 3LE, UK}

\author[0000-0002-0000-6977]{John D. Silverman}
\affiliation{Kavli Institute for the Physics and Mathematics of the Universe (WPI), The University of Tokyo, Kashiwa, Chiba 277-8583, Japan}
\affiliation{Department of Astronomy, School of Science, The University of Tokyo, 7-3-1 Hongo, Bunkyo, Tokyo 113-0033, Japan}

\author[0009-0003-4742-7060]{Takumi S. Tanaka}
\affiliation{Department of Astronomy, Graduate School of Science, The University of Tokyo, 7-3-1 Hongo, Bunkyo-ku, Tokyo, 113-0033, Japan}
\affiliation{Kavli Institute for the Physics and Mathematics of the Universe (WPI), The University of Tokyo Institutes for Advanced Study, The University of Tokyo, Kashiwa, Chiba 277-8583, Japan}
\affiliation{Center for Data-Driven Discovery, Kavli IPMU (WPI), UTIAS, The University of Tokyo, Kashiwa, Chiba 277-8583, Japan}

\author[0000-0002-3462-4175]{Si-Yue Yu}
\affiliation{Kavli Institute for the Physics and Mathematics of the Universe (WPI), The University of Tokyo, Kashiwa, Chiba 277-8583, Japan}
\affiliation{Department of Astronomy, School of Science, The University of Tokyo, 7-3-1 Hongo, Bunkyo, Tokyo 113-0033, Japan}

\author[0000-0003-3596-8794]{Hollis B. Akins}
\affiliation{The University of Texas at Austin, 2515 Speedway Blvd Stop C1400, Austin, TX 78712, USA}

\author[0000-0001-9610-7950]{Natalie Allen}
\affiliation{Cosmic Dawn Center (DAWN), Denmark} 
\affiliation{Niels Bohr Institute, University of Copenhagen, Jagtvej 128, DK-2200, Copenhagen, Denmark}

\author[0000-0002-7303-4397]{Olivier Ilbert}
\affiliation{Aix Marseille Univ, CNRS, CNES, LAM, Marseille, France  }

\author[0000-0002-6610-2048]{Anton M. Koekemoer}
\affiliation{Space Telescope Science Institute, 3700 San Martin Dr., Baltimore, MD 21218, USA}

\author[0000-0002-9489-7765]{Henry Joy McCracken}
\affiliation{Institut d’Astrophysique de Paris, UMR 7095, CNRS, and Sorbonne Université, 98 bis boulevard Arago, F-75014 Paris, France}

\author[0000-0003-2397-0360]{Louise Paquereau} 
\affiliation{Institut d’Astrophysique de Paris, UMR 7095, CNRS, and Sorbonne Université, 98 bis boulevard Arago, F-75014 Paris, France}

\author[0000-0002-4485-8549]{Jason Rhodes}
\affiliation{Jet Propulsion Laboratory, California Institute of Technology, 4800 Oak Grove Drive, Pasadena, CA 91001, USA}

\author[0000-0002-4271-0364]{Brant E. Robertson}
\affiliation{Department of Astronomy and Astrophysics, University of California, Santa Cruz, 1156 High Street, Santa Cruz, CA 95064, USA}

\author[0000-0002-7087-0701]{Marko Shuntov}
\affiliation{Cosmic Dawn Center (DAWN), Denmark} 
\affiliation{Niels Bohr Institute, University of Copenhagen, Jagtvej 128, DK-2200, Copenhagen, Denmark}

\author[0000-0003-3631-7176]{Sune Toft}
\affiliation{Cosmic Dawn Center (DAWN), Denmark} 
\affiliation{Niels Bohr Institute, University of Copenhagen, Jagtvej 128, DK-2200, Copenhagen, Denmark}

\begin{abstract}
We measure galaxy sizes from $2 < z < 10$ using COSMOS-Web, the largest-area JWST imaging survey to date, covering $\sim$0.54\,deg$^2$. 
We analyze the rest-frame optical ($\sim5000$ \AA) size evolution and its scaling relation with stellar mass ($R_e\propto M_*^\alpha$) for star-forming and quiescent galaxies. For star-forming galaxies, the slope $\alpha$ remains approximately 0.20 at $2 < z < 8$, showing no significant evolution over this redshift range. At higher redshifts, the slopes are  $-0.13 \pm 0.15$ and $0.37 \pm 0.36$ for $8 < z < 9$ and $9 < z < 10$, respectively. At fixed galaxy mass, the size evolution for star-forming galaxies follows $R_e \propto (1+z)^{-\beta}$, with $\beta = 1.21 \pm 0.05$. For quiescent galaxies, the slope is steeper $\alpha\sim 0.5$--$0.8$ at $2 < z < 5$, and $\beta=0.81\pm0.26$. We find that the size--mass relation is consistent between UV and optical at $z < 8$ for star-forming galaxies.  However, we observe a decrease in the slope from UV to optical at $z > 8$, with a tentative negative slope in the optical at $8 < z < 9$, suggesting a complex interplay between intrinsic galaxy properties and observational effects such as dust attenuation. We discuss the ratio between galaxies' half-light radius, and underlying halos' virial radius, $R_{vir}$, and find the median value of $R_e/R_{vir}=2.7\%$. The star formation rate surface density evolves as $\log\Sigma_\text{SFR} = (0.20\pm0.08)\,z+(-0.65\pm0.51)$, and the $\Sigma_\text{SFR}\textendash M_*$ relation remains flat at $2<z<10$. Lastly, we identify a threshold in stellar mass surface density $\log\Sigma_e\sim9.5\textendash10\, M_{\odot}/kpc^2$ marking the transition to compact, quenched galaxies from extended, star-forming progenitors. In summary, our findings show that the extensive COSMOS-Web dataset at $z > 3$ provides new insights into galaxy size and related properties in the rest-frame optical.
\end{abstract}

\keywords{galaxies: evolution – galaxies: fundamental parameters – galaxies: high-redshift – galaxies: structure }

\section{Introduction} \label{sec:intro}

In the standard $\Lambda$ cold dark matter ($\Lambda$CDM) cosmology, galaxy structure forms hierarchically. In the early universe, there are small density fluctuations in the overall matter distribution, where small dark matter halos form first and then aggregate into larger systems. Baryonic gas falls into the dark matter halos, undergoing processes such as cooling and condensation, leading to the formation of galaxies. Larger dark matter halos are expected to carry a larger amount of baryonic gas, hence corresponding to more massive galaxies \citep[for details, see review by][]{Wechsler2018}.  Additionally, both dark matter and diffuse gas acquire their angular momentum via tidal torques and galaxy mergers; therefore, the structure of galaxies follows the properties of dark matter to a certain degree.

Examining the evolution of galaxy size and its correlation with other physical parameters is vital to studying the fundamental physics of their growth. Over the past several decades, observations from both ground- and space-based telescopes have obtained fruitful results \citep{Shen2003, Williams2010ApJ, Mosleh2012,vdW2014, Shibuya2015, Yang2021, Kawinwanichakij2021, Yang2022a, Bouwens2022}. 
For example, in the local universe, \cite{Shen2003} utilized Sloan Digital Sky Survey data and found that the size distribution at a given luminosity is well described by a log-normal function at $z\lesssim0.3$. Later, the Hubble Space Telescope (HST) pushed our understanding to new frontiers. \cite{vdW2014} found that at the $0<z<3$ galaxy size evolves as $R_e\propto{(1+z)^{-0.75}}$ at a fixed stellar mass for star-forming galaxies, and the size–mass relation follows $R_e\propto M^{0.22}$. Their measurements were performed at rest-frame 5000 \AA.

At $z>3$, the wavelength range of Hubble limits the investigation of the evolution of galaxy size to the rest-frame ultra-violet (UV) \citep{Oesch2010, Shibuya2015, Bowler2017, Bouwens2017, Yang2022a}. For example, \citet{Shibuya2015} reports that the size of Lyman break galaxies evolves as $R_e\propto(1+z)^{-1.1}$ at $z=0$--$10$ at a given luminosity. The UV size-luminosity relation at these high redshifts is not only important for understanding galaxy formation and evolution but also has implications for the UV luminosity function and, hence, the properties of galaxies at the epoch of reionization  
\citep{Grazian2012, CurtisLake2016, Bouwens2017}.

The assembly history of quiescent galaxies is distinct from that of star-forming galaxies.
Furthermore, the size evolution at a given mass occurs faster, and the slope of the size--mass relation is steeper. For example, \cite{vdW2014} found that quiescent galaxy size evolves with redshift as $R_e\propto(1+z)^{-1.48}$, and its correlation with mass is $R_e\propto M^{0.75}$.
Unlike star-forming galaxies, the size growth of quiescent galaxies is primarily driven by galaxy mergers. Theoretically, two possible merger scenarios are each linked to the mass ratio of progenitor galaxies \citep[see derivation in][]{Naab2009}.  Minor mergers contribute size growth proportional to the increase in stellar mass, $\Delta R\propto\Delta M_{*}$, and major mergers between gas-poor galaxies of similar stellar mass produce size growth $\Delta R\propto\Delta M_{*}^2$.

Overall, the evolution of galaxy size in star-forming and quiescent galaxies, along with its scaling relation with mass and luminosity, have been well-characterized in the rest-frame optical using HST data at $z < 3$, and up to $z \sim 10$ for star-forming galaxies but in the rest-frame UV, (e.g., \citealt{Shibuya2015}). However, galaxies' size measured from the rest-frame UV is not a good tracer of the overall stellar mass. The UV light is emitted from young massive short-lived stars that may be distributed differently than the bulk of the stellar mass (older stars) and also suffer from significant dust extinction. Longer wavelength emission traces a galaxy's mass distribution more effectively because it is dominated by older, low-mass stars that constitute most of the galaxy's stellar mass. It is also less affected by dust extinction, providing a clearer view of the galaxy's true mass \citep{Forster2020}. Before the launch of the James Webb Space Telescope (JWST), simulations offered testable predictions for the rest-frame optical size of galaxies and its scaling relations with mass/luminosity at high redshifts ($z>3$), as well as size as a function of wavelength \citep{Ma2018, Wu2020, Popping2021, Roper2022, Marshall2022, Shen2024}, and we are able now to test that.

JWST is now revolutionizing our understanding of galaxies by providing broad wavelength coverage, unprecedented depth, and high spatial resolution across a broad wavelength range. At $z>3$, the NIRCam instrument (e.g., with the F444W filter) enables us to probe the rest-frame optical size of galaxies out to $z\sim7$ at a similar angular resolution to that provided by HST WFC3-IR in the rest-frame UV (i.e., F160W).  In addition, multiple NIRCam filters sampling the rest-frame optical (e.g., covering the 4000 \AA\,break) benefit the determination of galaxies' stellar populations and their properties, such as stellar mass. Many recent JWST studies focus on galaxy structures in the rest-frame optical, examining their relationship with luminosity or stellar mass and comparing these findings with those obtained from rest-frame UV observations or simulations. For instance, \citet{Yang2022b} explored galaxy size and its wavelength-dependent correlation with luminosity at $z>7$, finding that the median size in the rest-frame optical is compatible but slightly larger than in the UV.  
Several studies have investigated the evolution of the size--mass relation of star-forming galaxies at $z>3$ \citep{Morishita2024, Ormerod2024, Ward2024, Allen2024, Miller2024}. The number of observed quiescent galaxies at $z>3$ has also increased significantly, enabling more robust statistical analyses \citep{Ito2024, Wright2024}. However, these studies are based on relatively small sample sizes. To enable an unbiased statistical analysis, a larger sample of both star-forming and quiescent galaxies is required.

The COSMOS-Web survey, the largest survey selected in JWST Cycle 1 (PIs: J. Kartaltepe \& C. Casey, \citealt{Casey2023}) covering $\sim0.6$\,deg$^2$ provides an extensive dataset for this study.  The primary goals of this work are to 1) investigate the size--mass relation of both star-forming and quiescent galaxies at $z=2$--$10$ and how their properties vary as a function of wavelength, 
2) validate the predictions made by cosmological simulations, and 3) gain insights into galaxy growth and quenching mechanisms through the evolution of the star formation rate surface density.

This paper is organized as follows.  The data and sample selection for this study are described in Section \ref{sec:data}. In Section \ref{sec:method}, we present the methodology of our morphological measurements and size--mass relation fitting. In Section \ref{sec:results}, we report the results of the size--mass relation and its evolution with redshift for star-forming and quiescent galaxies, as well as its variation as a function of wavelength in Section~\ref{sec:size-wavelength}. 
We also study the size distribution histogram in different mass bins as a complementary analysis of the size--mass relation in Section \ref{sec:size-dis}. Lastly, we discuss and summarize our results in Section \ref{sec:dis} and \ref{sec:summary}, respectively. 
Througout this work, we adopt the standard $\Lambda$CDM cosmology with $\Omega_{\rm m}=0.3$, $\Omega_{\Lambda}=0.7$, and H$_0$=70 km s$^{-1}$ Mpc$^{-1}$, AB system magnitudes \citep{Oke1983}, and a \citet{Chabrier2003} initial mass function for computing stellar masses.

\section{DATA} \label{sec:data}

\subsection{The COSMOS-Web survey} \label{sec:style}
The COMOS-Web survey is the largest survey selected for observations during JWST Cycle 1 and consists of 0.54 deg$^2$ Near-Infrared Camera (NIRcam) observations in four filters (F115W, F150W, F277W and F444W) and 0.19 deg$^2$ Mid-Infrared Imager (MIRI) observations in F770W \citep[PIs: Kartaltepe \& Casey,][]{Casey2023}. The depth of the NIRCam data is measured to be 26.6--27.3 mag (F115W), 26.9--27.7 mag (F150W), 27.5--28.2 mag (F277W), and 27.5--28.2 mag (F444W) for 5$\sigma$ point sources calculated within 0.15 arcsecs radius apertures. The depths of the MIRI observations are 25.33--25.98 mag calculated within 0.3 arcsecs radius apertures. The data reduction is summarized by (\citealt{Franco2024} and full details will be described by Franco et al. (in prep) and Harish et al. (in prep.).

The COSMOS-Web data were obtained during three observing windows, January 2023, April 2023, and December 2023/January 2024 with a handful of pointings observed during April/May 2024. 
The final NIRcam mosaics \footnote{https://cosmos.astro.caltech.edu/page/cosmosweb} are created with two different pixel scales, 30\,mas and 60\,mas. In this paper, we use the 30\,mas pixel scale mosaic.

\subsection{COSMOS-Web Photometric Catalog} \label{sec:style}
Details of the multi-wavelength photometry extraction and catalog construction will be fully described by Shuntov et al. (in prep).
Here, we summarize key relevant features.

\subsubsection{SE++ Photometry}
Shuntov et al. (in prep) performed source extraction across 33 ground- and space-based filters using a multi-band model-fitting approach with \textsc{SourceXtractor++} \citep[SE++][]{Bertin2020, Kummel2020}, which is an updated version of the widely used \texttt{Source Extractor} package \citep{Bertin1996}. SE++ fits S\'ersic parametric models to extract source photometry for all detected sources across  available filters. For each source, the structural parameters are kept the same for all filters, thus, the structural parameters correspond to the averaged morphology over a wide wavelength range. For the purpose of this work and also to maximize the scientific application of the catalog, i.e., investigating galaxy structure as a function of wavelength, we provide an alternative morphology catalog (Yang et al. in prep) that models the galaxies via S\'ersic model individually in four NIRCam bands, as described in Section~\ref{sec:method}.

\subsubsection{SED fitting}
The template spectral energy distribution (SED) fitting code \textsc{LePHARE} \citep{Arnouts02, Ilbert06} is used to derive photometric redshifts and physical parameters from the SE++ model based photometry for each source in the full catalog \citep{Shuntov2024}. The template fitting is based on \citet{BC03} models with diverse star formation histories, ages, and dust attenuation curves \citep{Calzetti00, Arnouts2013, Salim18}. Emission lines and intergalactic medium absorption are modeled following \citet{Saito20, Schaerer09}, and \citet{Madau95}, respectively.  \textsc{LePHARE} provides the probability density function of the photometric redshift for each source, and the median value is adopted in this work. The physical parameters (e.g., stellar mass, SFR, etc.) are then calculated using this fixed redshift.

Based on a comparison with available spectroscopic redshifts in the field \citep{Khostovan2025}, the photometric redshift computed via \textsc{LePHARE} have a high level of confidence. 
The scatter $\sigma_{\text{NMAD}}$ is approximately 0.013 for galaxies with $m_\text{F444}<25$, 
where $\sigma_{\text{NMAD}}=1.48\times \text{median}\left[(\left|\Delta z - \text{median}(\Delta z)\right|)/(1 + z_{\text{spec}})\right]; \quad \Delta z=z_{\text{phot}} - z_{\text{spec}}$, see Table 1 in \citet{Shuntov2024}. Additional assessment of stellar mass estimates is delivered by comparison with results measured by \textsc{Cigale} \citep{Boquien19}, employing non-parametric star formation histories modeling and alternative dust attenuation laws. Detailed comparisons between \textsc{LePHARE} and \textsc{Cigale} are also provided by \citet{Shuntov2024}.
In this work, we adopt the stellar mass obtained by \textsc{LePHARE}.

\subsection{Sample selection}

\begin{figure*}
\includegraphics[width=2\columnwidth]{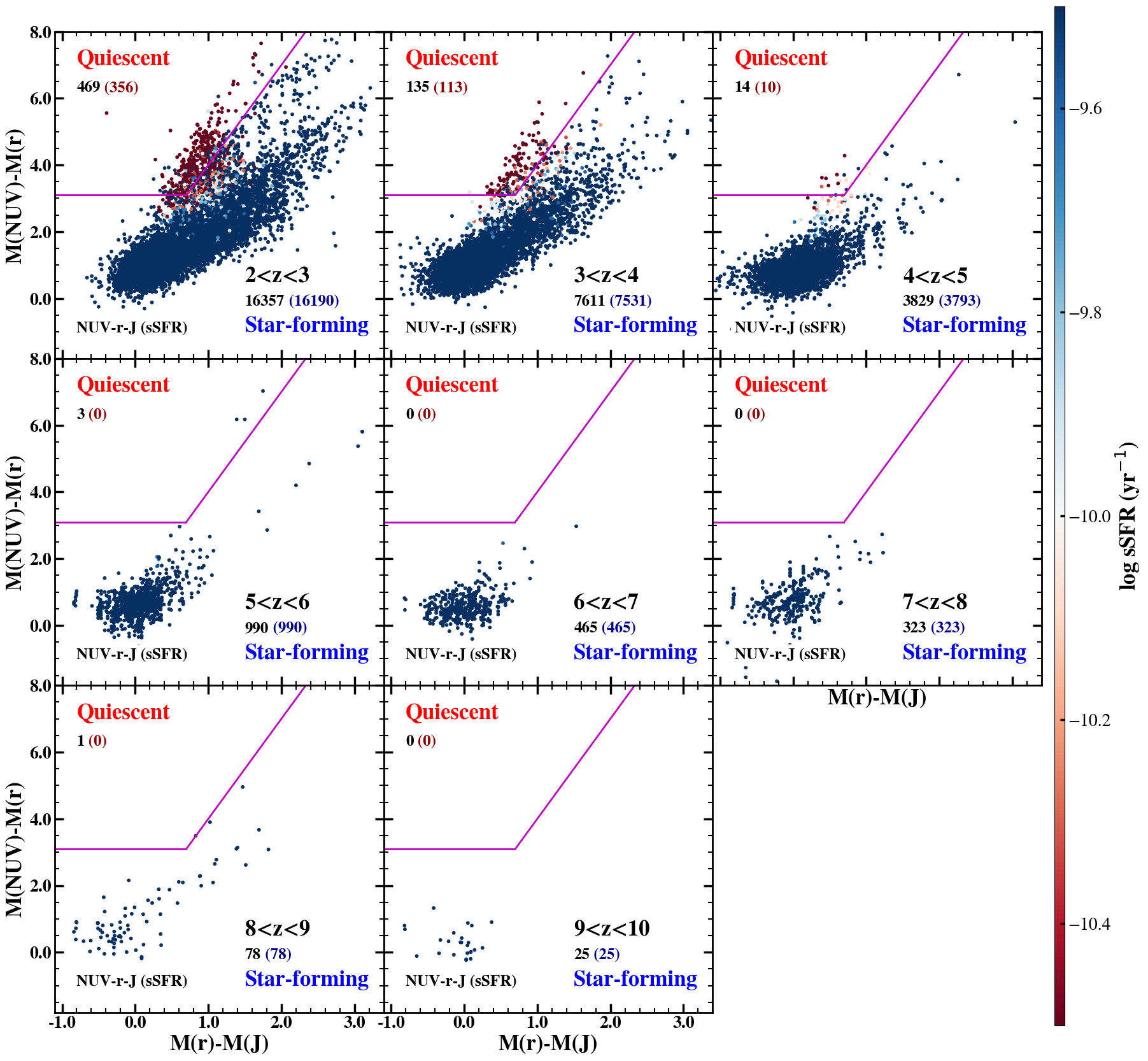}
\caption{Rest-frame NUV-r-J color-color diagram for galaxies at $2<z<10$.
Data points are color-coded according to their specific star formation rate (sSFR). The magenta line indicates the NUV-r-J criteria (see Eq.~\ref{eq:nuvrj}), separating quiescent galaxies (top left) from star-forming galaxies (bottom right). Additionally, galaxies are required to meet the sSFR criterion $\log \text{(sSFR)}<0.2/t_\text{obs}$, where $t_\text{obs}$ is the observed Universe age. In each panel, the numbers of quiescent and star-forming galaxies classified using the color-color diagram are labeled in black, while those further classified with the additional sSFR criterion are labeled in blue and red inside the bracket.}
\label{img:nuvrj}
\end{figure*} 

We initially selected our sample at $2<z<10$ in 8 redshift bins with bin width $\Delta z=1$ from the parent COSMOS-Web SE++ galaxy catalog. Following \citet{Shuntov2024}, we apply the bright star mask to remove sources near bright stars, and a F444W magnitude cut ($m_{F444W}=27.5$) to ensure the flux in the rest-frame optical is robustly measured. We adopt the median value of the redshift probability density function as the photometric redshift in this work, and remove sources with poorly constrained photo-z, specifically those with a 68\% confidence level width greater than the width of our redshift bins. Additionally, we remove stars and brown dwarfs using the $\chi^2_\text{star}$ of the star/brown dwarf SED templates fit by \textsc{LePHARE}.

We also exclude potential AGN to avoid contamination in the measurement of galaxy physical properties. We remove sources with X-ray counterparts by cross-matching within 1 arcsec with the catalog of \citet{Civano2016} or those listed in the Little Red Dots (LRDs) catalog of \cite{Akins2024}. LRDs are very red compact sources identified at high redshift that potentially host strong AGN (e.g., \citealt{Kocevski23, Matthee24}. We identify further AGN candidates by requiring that $\chi^2_\text{AGN} < \chi^2_\text{gal}$ and that the size is smaller than 60\,mas (corresponding to half the FWHM of the F277W PSF).

To ensure that our galaxy sample is above the completeness limit at all redshifts, we select galaxies with stellar mass $\log M_{*}/M_{\odot}>9$. This mass cut is chosen because, as demonstrated in \citet{Shuntov2024}, the COSMOS-Web survey achieves $95\%$ stellar mass completeness down to $\log M_{*}/M_{\odot}=7.5\textendash8.8$ at $0.2<z<12$. The galaxies with $\log M_{*}/M_{\odot}>9$ are well above the completeness threshold for the entire redshift range of $2<z<10$. With this mass cut, we initially select 30,300 galaxies from the catalog.

One of our main goals is to study the size--mass relation of star-forming galaxies (SFGs) and quiescent galaxies (QGs). The next step is to separate galaxies into those two classes. Following the approach of \cite{Ilbert2013}, we utilize the NUV-r-J color diagram  to separate the sample into quiescent and star-forming galaxies. 
Galaxies with criteria,
\begin{equation}
\centering
\begin{aligned}
M_{NUV} - M_r > 3(M_r - M_J)+1 \\
\text{and}\,\,\, M_{NUV} - M_r > 3.1 
\label{eq:nuvrj}
\end{aligned}
\end{equation}
are classified as quiescent galaxies. Figure~\ref{img:nuvrj} shows the distribution of our sample in NUV-r-J space and color-coded by their specific star formation rate (sSFR). The magenta lines in each panel mark the division, with galaxies in the top left corner considered to be quiescent and the rest classified as star-forming.

The sSFR is also commonly used to separate quiescent and star-forming galaxies.  Quiescent galaxies are often defined as having $\log \text{(sSFR)}<0.2/t_\text{obs}$, where $t_\text{obs}$ is the age of the Universe at the corresponding redshift \citep{Pacifici2016, Carnall2023}.  Galaxy counts classified by the color-color criteria and with the sSFR criteria are also labeled in Figure~\ref{img:nuvrj}. For example, at $2<z<3$,  we identify 469 galaxies as quiescent in the NUV-r-J diagram, with 356 (76\%) of these also satisfying the threshold $\log\text{(sSFR)}/yr^{-1}<-10.1$. Within a redshift bin, we adopt the same threshold that is calculated via the center redshift value. In our work, to select a conservative sample of quiescent galaxies, we require that galaxies pass both the color-color and sSFR criteria. For star-forming galaxies, we also impose the condition that $\log\text{(sSFR)}>0.2/t_\text{obs}$. Galaxies that pass the color-color criteria but do not satisfy the sSFR criterion are excluded. Our sample now comprises 29,874 galaxies, including 29,395 SFGs and 479 QGs.

The sample size of quiescent galaxies can vary based on the selection method—whether using NUV-r-J, the traditional UVJ, or methods utilizing NIRCam filters \citep{Long2024}, which could introduce potential systematic uncertainties. NUV-r-J is a modified version of the UVJ diagram proposed by \cite{Williams2009}. Our results show that quiescent galaxies selected using the NUV-r-J criteria align more closely with those selected by sSFR, in contrast to UVJ, where only about 50\% meet the sSFR threshold at $2<z<3$. Hence, we adopt the NUV-r-J method in this work.

\section{Size Measurement} \label{sec:method}
To measure the size of the galaxies in our sample, we use 2D S\'ersic model to fit the surface brightness profile of galaxies \citep{Sersic1968},
\begin{equation}
I(r) = I_0 \exp \left[ -b_n \left(\frac{r}{r_e}\right)^{\frac{1}{n_{\text{s\'ersic}}}} - 1\right],
\end{equation} 
where $I_0$ is the surface brightness amplitude at the half-light radius $r_{\text{e}}$, $n_{\text{s\'ersic}}$ is the S\'ersic index, and $b_n$ is a parameter dependent on $n_{\text{s\'ersic}}$. $r=\sqrt{x^2+y^2/q^2}$, where $(x,y)$ is the coordinates of image and $q$ is the axis ratio. We utilize the \texttt{python} software package \texttt{Galight} \citep{Ding2020} to perform the surface brightness profile fitting \footnote{https://github.com/dartoon/galight}, which inherits the image modeling capabilities of \texttt{Lenstronomy} \citep{Birrer2021}.

The detailed process of size measurement and uncertainty assessment will be described by Yang et al. (in preparation), but here we provide a summary. Before fitting the surface brightness profile, it is necessary to determine the appropriate image cutout size, which should balance computational efficiency while being large enough to encompass the entire galaxy. We use a cutout radius of 5 times the size of the galaxy as measured in the parent SE++ catalog, with a minimum radius of 30 pixels and a maximum radius of 200 pixels. A corresponding noise map, i.e., the ERR extension in the i2d fits file,  of the same size is also created. The noise map is computed as the square root of the total variance, which includes contributions from background noise, readout noise, and Poisson noise. In some cases, the image cutout may include emissions from nearby sources. We either mask out or model the contaminating sources using additional S\'ersic models, depending on their proximity to the central galaxy. For the S\'ersic parameters, we set the range of $r_{\text{e}}$ to be between 0.01 arcseconds and the size of the cutout image, $n_{\text{s\'ersic}}$ between 0.3 and 9, and the axis ratio $q$ between 0.1 and 1. Disk-like galaxies typically have $n_{\text{s\'ersic}}\sim1$, while some diffuse galaxies can have even lower values. To accommodate these cases while preventing unphysically shallow profiles, we set the lower boundary of $n_{\text{s\'ersic}}$ to 0.3.  Elliptical galaxies typically have $n_{\text{s\'ersic}}\sim4$, and some highly concentrated systems such as brightest cluster galaxies can have extremely high values, therefore we set the upper limit to 9. For the axis ratio $q$, edge-on disk galaxies have $q\sim0.2\textendash0.3$, while $q=1$ indicates a perfect circular system. To ensure reasonable fits, we set $q$ to 0.1--1. The point spread function (PSF) is also crucial for accurate surface brightness profile fitting. The PSF for each filter used in this work is constructed using  \texttt{PSFEx}\footnote{\url{https://github.com/astromatic/psfex}} \citep{Bertin2011}, see details by Shuntov et al., in preparation.

\begin{figure}
\includegraphics[width=\columnwidth]{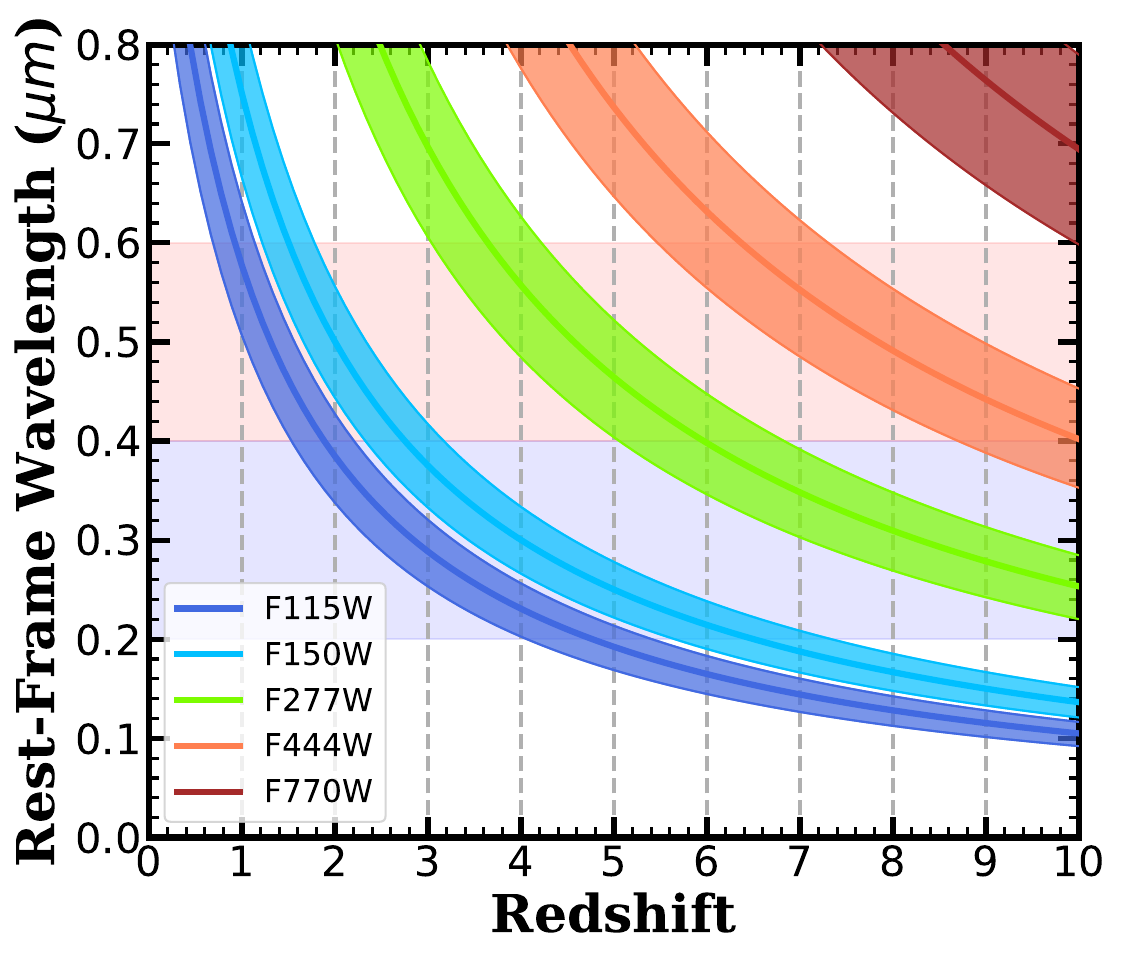}
\caption{The rest-frame wavelength probed by each of the COSMOS-Web filters as a function of redshift. The shaded background purple (red) regions indicate the rest-frame UV (optical) range, i.e., 2000--4000 \AA\, (4000--6000 \AA) for UV (optical). For example, at $6<z<10$, the F277W (F444W) filter probes the rest-frame UV (optical) properties of galaxies.}
\label{img:filters}
\end{figure} 

One of the main goals of this work is to study size-related properties as a function of wavelength. To measure the sizes of galaxies in the rest-frame optical and UV, we need to identify the corresponding NIRCam images observed in the appropriate filters. We select the rest-frame wavelength range 2000--4000 \AA \,(4000--6000 \AA) to represent the UV (optical). As shown in Figure~\ref{img:filters}, the appropriate filters for each rest-frame wavelength range vary between different redshift bins. For example, we select F277W (F444W) to measure the rest-frame UV (optical) at $6<z<10$.

We apply the above strategy to fit each of the galaxies in the rest-frame UV and optical. We discard the results if any of the parameters hit the boundaries of the provided ranges, or have a catastrophic fitting performance, i.e., reduced $\chi^2>15$. After removing these sources, our final galaxy sample consists of 28,274 galaxies in total, including 27,861 SFGs and 413 QGs.

%%%%%%%%%%%%%%%%%%%%%%%%%%%%%%%%%%%%%%
\section{results}\label{sec:results}

%%%%%%%%%%%%%%%%%%%%%%%%%%%%%%%%%%%%%%
\subsection{size--mass Distribution at $2<z<10$}

%figure
\begin{figure*}
\includegraphics[width=2\columnwidth]{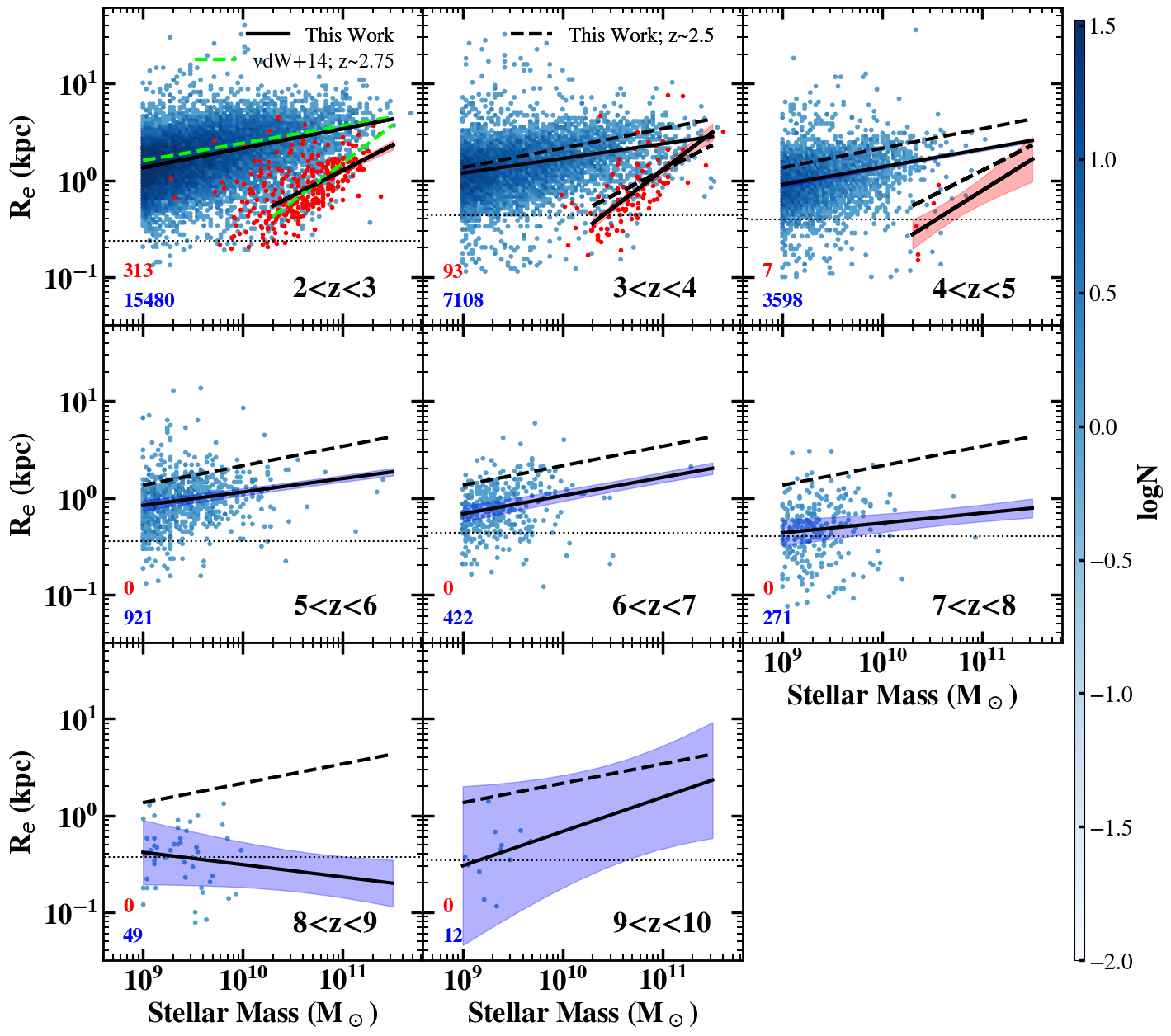}
\caption{The rest-frame optical size--mass distribution of star-forming (blue) and quiescent (red) galaxies presented in redshift bins at $2<z<10$. The black solid lines indicate the best-fit size--mass relation of star-forming and quiescent galaxies, (see also Table~\ref{table:size--mass}). The thick black and lime dashed lines show the fitting results at $2<z<3$ in this work and $z\sim2.75$ obtained by \cite{vdW2014}, respectively. The horizontal black dotted line in each panel indicates the PSF size (FWHM/2) of the filter that is closest to the rest-frame optical in each redshift bin. The numbers of galaxy data points are labeled at bottom left in each redshift bin.
}
\label{img:optical-sm}
\end{figure*}

We present the size--mass distribution of galaxies in the rest-frame optical as a function of redshift in Figure~\ref{img:optical-sm}. Intuitively, more massive galaxies tend to have larger sizes across most redshift bins, a trend observed for both star-forming (blue) and quiescent galaxies (red). However, at a given stellar mass, the scatter in size is substantial. On average, quiescent galaxies are smaller than their star-forming counterparts. Following earlier work by \citet{Shen2003} and \citet{vdW2014}, in each redshift bin we fit a linear relation between the logarithm of stellar mass ($M_*$) and effective radius ($R_e$)
%equation
\begin{equation} \label{re}
\log\frac{R_e}{kpc}= \log(A) + \alpha \log\frac{M_{*}}{5\times10^{10}M_{\sun}},
\end{equation}
where $\log(A)$ is the y-intercept at the characteristic stellar mass $5\times10^{10} M_{\odot}$ and $\alpha$ is the slope. At a given stellar mass, the effective radius obeys a log-normal distribution $\mathcal{N}(\log R_e (M_*), \sigma_{\rm \log R_e}^2)$, where $\sigma_{\rm \log R_e}$ is the intrinsic scatter.

We then use a standard Bayesian approach to derive the posterior distributions for the parameters, the y-intercept $\log(A)$, slope $\alpha$, and $\sigma_{\rm \log R_e}$. Our analysis incorporates both statistical uncertainties from the Markov Chain Monte Carlo (MCMC) fitting process, and systematic uncertainties arising from variations in initial settings, and from different software measurements, i.e., SE++ and \texttt{Galight} (see Yang et al., in preparation for details). The total uncertainty in $\log R_e$ is $\sim0.1\textendash0.2$ dex, depending on the filter wavelength and redshift, with systematic uncertainties dominating. 

Additionally, following \cite{vdW2014}, we account for size uncertainties related to the stellar mass measurements. The median $\log M_*$ uncertainty is $0.09\textendash0.24$ dex at $z=2.5\textendash9.5$ and we use $\alpha=0.7$ and $0.2$ to convert these into size uncertainties for quiescent and star-forming galaxies, respectively. It has been noted that the scaling relation for quiescent galaxies flattens for stellar masses below $2\times10^{10} M_{\odot}$ \citep{vdW2014, Cutler2024}. Therefore, we restrict our fitting to quiescent galaxies above this mass threshold. Figure~\ref{img:optical-sm} displays the fitting results (see also Table~\ref{table:size--mass}) with solid black lines and shaded regions representing the best-fit relations and their $1\sigma$ uncertainty range. The horizontal black dotted lines indicate the PSF size (FWHM/2), and the fitting results at $2<z<3$  are shown as dashed lines in each panel for reference. For comparison with previous HST results, we also include the best-fit line from \citet{vdW2014} at $z\sim2.75$, as shown by the lime dashed line.

%%%%%%%%%%%%%%%%%%%%%%%%%%%%%%%%%%%%%%
\subsection{Size evolution}
%figure
\begin{figure*}[t]
\includegraphics[width=1.8\columnwidth]{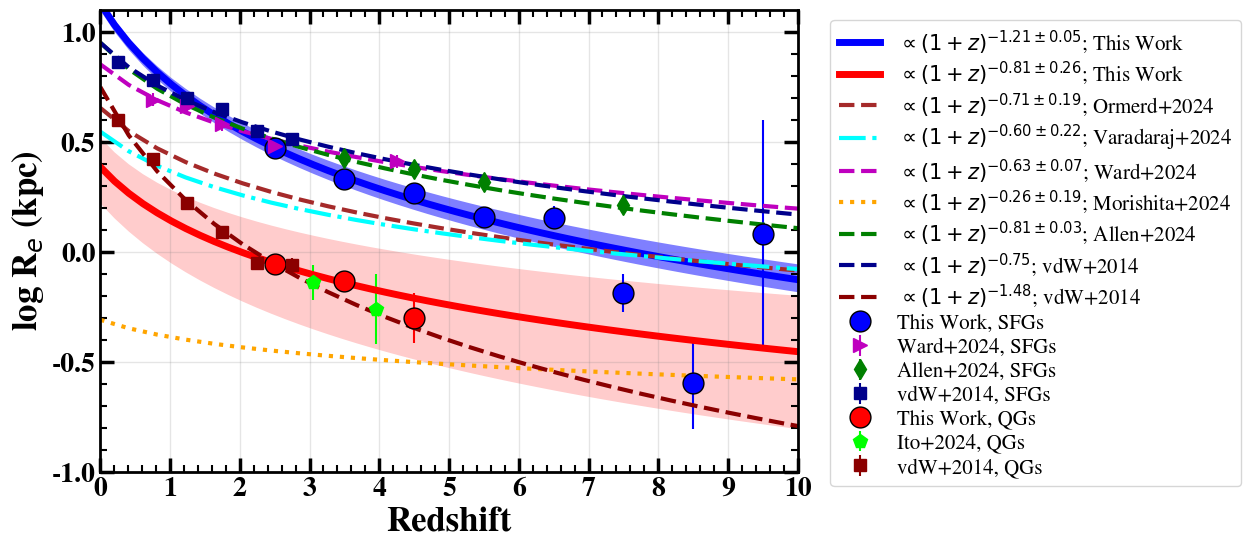}
\includegraphics[width=\columnwidth]{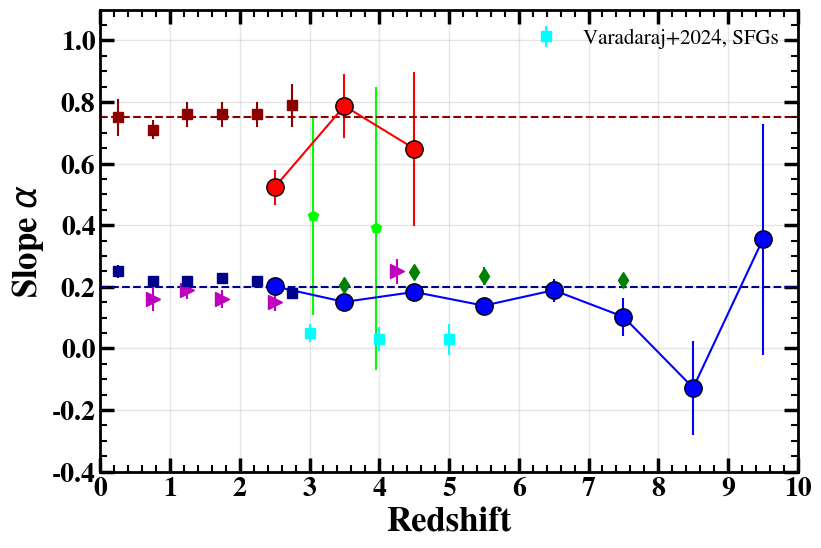}
\includegraphics[width=\columnwidth]{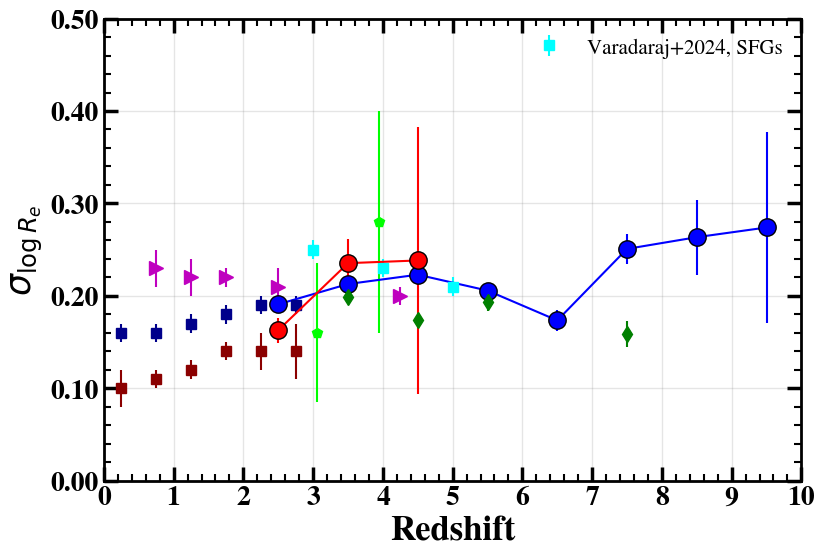}

\caption{(Top) The evolution of rest-frame optical sizes for star-forming galaxies (SFGs, blue circles) and quiescent galaxies (QGs, red circles) at a fixed stellar mass $5\times10^{10}M_{\odot}$ at $2<z<10$, with comparisons to previous studies. The fitting results for SFGs and QGs are shown by the blue and red solid lines, respectively, corresponding to $R_e \propto (1+z)^{-1.21\pm0.05}$ and $R_e \propto (1+z)^{-0.81\pm0.26}$, with the shaded regions representing the $1\sigma$ uncertainties. Utilizing recent JWST data, the results of SFGs from \cite{Ormerod2024}, \citet{Varadaraj2024}, \cite{Ward2024}, \citet{Morishita2024} and \citet{Allen2024} are illustrated by brown, cyan, magenta, orange and green colors, respectively. Lime pentagons display results for QGs \citep{Ito2024}. The dark blue and red represent HST results of SFGs and QGs, from \cite{vdW2014}. (Bottom) the evolution of the slope $\alpha$ (left) and intrinsic scatter $\sigma_{\log \text{R$_e$}}$ (right) with colors the same as top panel.  }
\label{img:optical-size-z}
\end{figure*} 

The evolution of the rest-frame optical size at a given mass, characterized by the y-intercept $\log(A)$, for SFGs and QGs is shown in the top panel of Figure \ref{img:optical-size-z}. The sizes of both SFGs and QGs decrease toward higher redshifts. We parameterize this size evolution as a function of the cosmological scale factor $(1+z)$.
We find that the size evolution of SFGs follows $R_e\propto(1+z)^{-1.21\pm0.05}$, with the blue solid line showing the fitting result, with the shaded region indicating the $1\sigma$ uncertainty.  We compare our results to those from recent JWST studies \citep{Ormerod2024, Ward2024, Varadaraj2024, Morishita2024, Allen2024} and a previous HST study \citep{vdW2014}. Our fitting result suggests a faster size evolution than previous results, while these previous works found flatter slopes ranging from $-0.26$ to $-0.81$. At $2<z<3$, where measurements in the rest-frame optical from both HST and JWST are available, our measured size ($\log R_e\text{/kpc}\sim0.47$) is consistent with other JWST results (e.g., \citealt{Ward2024} report $\log R_e\text{/kpc}\sim0.48$) but slightly smaller than HST results (e.g., \citealt{vdW2014} obtained $\log R_e\text{/kpc}\sim0.51\pm0.01$ at $z\sim2.75$). We speculate that this subtle offset arises from the higher resolution of NIRCam F150W imaging compared to the HST H$_{F160W}$ band, which covers a similar wavelength range.

At $z>3$, our sizes inferred from the size--mass scaling relation are smaller than those reported by \citet{Ward2024} and \citet{Allen2024}, resulting in faster size evolution. This discrepancy possibly arises from several factors. First, both studies have limited sample sizes and use broader redshift bins at high redshifts. Although \citet{Allen2024} have sufficient data at $5<z<6$, the number of high-mass galaxies ($\log (M_*/M_{\odot})>9.5$) is limited. Second, differences in slope determination for the size--mass relation can lead to variations in inferred sizes (see the bottom left panel of Figure~\ref{img:optical-size-z}). For example, at $7<z<8$, we obtain the slope $\alpha=0.10$ $\log(A)$ which is very flat compared to the slope obtained in \citet{Allen2024}. Furthermore, the COSMOS-Web survey is conducted with four NIRCam filters, while other JWST surveys have coverage with more filters (e.g., \citealt{Ward2024} uses imaging in seven NIRCam filters from CEERS: \citealt{Finkelstein2025}), enabling those studies to select bands more closely aligned with rest-frame 5000 \AA\ for analysis. However, since galaxy sizes do not vary significantly over this narrow wavelength range, especially at $z>2$, we do not expect this to be the primary source of the observed differences.

The size evolution trends discussed by \citet{Ward2024} and \citet{Allen2024} are derived for galaxies at a characteristic stellar mass of $5\times10^{10}M_\odot$, enabling a fair comparison across studies. However, some other studies use different stellar mass ranges or quoted sizes when deriving size evolution. For example, \cite{Varadaraj2024} analyzed the evolution of the average size with $\log (M_*/M_{\odot})>9$ at $z=3\textendash5$ and they found $(1+z)^{-0.60\pm0.22}$ across sample. \cite{Ormerod2024} analyzed $\sim$1400 galaxies with stellar mass $\log (M_*/M_{\odot})>9.5$ at $0.5<z<8$, reporting an average rest-frame optical size evolution, $(1+z)^{-0.71\pm0.19}$. Additionally, \cite{Morishita2024} obtained a much slower rate of size evolution, $(1+z)^{-0.26\pm0.19}$, using $\sim$400 galaxies with stellar mass as low as $\log (M_*/M_{\odot})\sim7$ .

While the size evolution of SFGs provides insights into the growth of actively forming systems,  the trends for quiescent galaxies offer a complementary perspective on the structural changes in galaxies that have ceased significant star formation. For QGs, the red data points in Figure \ref{img:optical-size-z} show the y-intercepts $\log(A)$ at $2<z<5$. The red solid line represents the fitting result, $R_e\propto(1+z)^{-0.81\pm0.26}$. Due to the fact that there are fewer QG samples and we only have results in three redshift bins, the $1\sigma$ uncertainty for the slope of the size evolution is relatively large, as shown by the shaded region. At $2<z<3$, our results are consistent with recent JWST results from \cite{Ito2024} (lime pentagons), as well as HST results from \cite{vdW2014} (dark red squares). At $z>3$, our data points are also consistent with the results extrapolated from the size evolution of $(1+z)^{-1.48}$ derived from \cite{vdW2014}.

%%%%%%%%%%%%%%%%%%%%%%%%%%%%%%%%%%%%%%
\subsection{Evolution of the Slope of the Size--Mass Relation}
In the bottom left panel of Figure~\ref{img:optical-size-z}, we show the evolution of the slope, $\alpha$, for both SFGs and QGs. For SFGs $\alpha$ ranges from 0.10 to 0.20 at $2<z<8$, showing no significant variation across these redshift bins. At the highest two redshift bins ($8<z<9$ and $9<z<10$), the slopes are $-0.13\pm0.15$ and $0.37\pm0.36$, respectively. These values are poorly constrained due to the limited number of data points and large scatter. We compare our findings with other works. At $2<z<3$, our results are consistent with \citet{vdW2014}, who reported an average slope of 0.22 at $z<3$ (dark blue squares). Their measured slope is $0.18\pm0.02$ at $z\sim2.75$ agrees well with our result of $0.20\pm0.01$ at $z=2.5$, as well as with the slope reported by \citet{Ward2024} (magenta triangles). At higher redshifts ($3<z<8$), we observe slightly flatter slopes in some bins ($3<z<4$, $5<z<6$, and $7<z<8$) with the flattest slope being $\alpha = 0.10\pm0.06$ at $7<z<8$; however, this is still consistent with 0.20 within $2\sigma$. \citet{Allen2024} (green diamonds) also found a relatively constant slope of $\sim0.2$ over $3<z<8$. However, much flatter slopes at $3<z<5$ have been reported by \citet{Varadaraj2024} (cyan squares), though their sample size is smaller.

For QGs, the slopes are $0.52\pm0.06$, $0.78\pm0.11$, and $0.64\pm0.26$ at three redshift bins between $z=2$ to $z=5$, respectively. At $2<z<3$, our result aligns with the results of \citet{Ito2024} (lime pentagons), who recently analyzed two dozen quiescent galaxies at $2.8<z<4.6$ and found slopes of $\alpha \sim 0.4\textendash0.55$. However, our result is slightly shallower than the average value of $\alpha \sim 0.75$ (dark red dashed line) reported by \citet{vdW2014} at lower redshift. One possible explanation for this difference is the selection criteria for QGs. We use the NUV-r-J diagram with an additional sSFR requirement, while \citet{vdW2014} adopted the UVJ diagram. Other contributing factors include the increasing fraction of green valley galaxies and changes in the stellar mass distribution (see the discussion of \citet{Ito2024} for more details). At $3<z<4$ and $4<z<5$, our result is consistent with that of \citet{vdW2014}, but the uncertainties are relatively large due to the limited sample size.

\subsection{Evolution of Intrinsic Scatter}

The evolution of intrinsic scatter, $\sigma_{\log R_e}$, is shown in the bottom right panel of Figure~\ref{img:optical-size-z}. For SFGs, $\sigma_{\log R_e}$ remains around $\sim0.2$ dex at $2<z<7$ and increases to $\sim0.25$ dex at $7<z<10$ with large uncertainties. Our result at $2<z<3$ is consistent with those of \citet{vdW2014} and \citet{Ward2024}. \citet{Varadaraj2024} measured intrinsic scatter values of $0.21-0.25$ dex in F356W for galaxies at $z=3-5$ (cyan points), which generally agree with our results, though their values are slightly higher at $z\sim3$. Similarly, \citet{Allen2024} found $\sigma_{\log R_e} \lesssim 0.2$ dex at $3<z<8$, which is also in general agreement with our results but is slightly lower at $z\sim7.5$. For QGs, $\sigma_{\log R_e}$ is $0.16\pm0.01$ dex at $2<z<3$ and smaller than that of SFGs, which is consistent with the results reported by \citet{vdW2014}. It increases to $0.24\pm0.03$ and $0.25\pm0.25$ dex at higher redshifts where the values are comparable to those for SFGs. Large intrinsic scatters have also been reported by \citet{Ito2024} at $3<z<5$.

The properties of intrinsic scatter are closely related to the spin of dark matter halos. If we assume SFGs follow the characteristics of disk galaxies, their growth is expected to align with the evolution of their dark matter halos. Consequently, the intrinsic scatter in galaxy sizes should resemble the scatter in dark matter halo spin parameters. \citet{Maccio2008} presented the distribution of halo spin parameters, $\log \lambda$, finding a $1\sigma$ scatter of $\sim0.24$ dex, which is independent of halo mass and the choice of cosmological models. The distribution of spin parameters also does not change significantly as a function of redshift \citep{Munoz2011}. Our findings show a consistent size scatter across a wide redshift range ($2<z<7$), albeit with slightly lower values than the spin parameter. This suggests that the sizes of disk galaxies are generally determined by the properties of their underlying dark matter halos. At $z>7$, the $\sigma_{\log R_e}$ increase, and it is possibly because of the higher galaxy interaction rate at an earlier time. For QGs, the intrinsic scatter at higher redshifts is $\sim0.25$ dex. We speculate that this similarity in scatter with SFGs implies some shared characteristics with disk galaxies (see more discussion in Section~\ref{sec:lshape}), possibly reflecting a common dependence on dark matter halo properties.

%table2
\begin{table*}
	\centering
	\caption{The fitting results for the size--mass distribution, expressed as $\log R_e/(kpc)= \log (A) + \alpha \log M_{*}(5\times10^{10}M_{\odot})$,
    are recorded for star-forming and quiescent galaxies at $2<z<10$ in both the rest-frame optical and UV.}
		\begin{tabular}{lccccccc} 
		\hline

		\hline
		 &   & Star-forming galaxies & &  & & Quiescent galaxies &    \\ 
   %\cline{2-4}\cline{6-7}			
 $z$ & $\alpha$ &$\log(A)$ & $\sigma$ log($R\textsubscript{eff}$) & &$\alpha$ &$\log(A)$ &  $\sigma$ log($R\textsubscript{eff}$) \\		
\hline		
&&&&Optical&&&\\
\hline
2.5 & 0.20 $\pm$ 0.01 & 0.47 $\pm$ 0.01 & 0.19 $\pm$ 0.01&& 0.52 $\pm$ 0.06 & -0.05 $\pm$ 0.01 & 0.16 $\pm$ 0.01\\ 
3.5 & 0.15 $\pm$ 0.01 & 0.33 $\pm$ 0.01 & 0.21 $\pm$ 0.01&& 0.78 $\pm$ 0.11 & -0.13 $\pm$ 0.03 & 0.24 $\pm$ 0.03\\
4.5 & 0.18 $\pm$ 0.01 & 0.27 $\pm$ 0.02 & 0.22 $\pm$ 0.01&& 0.64 $\pm$ 0.26 & -0.31 $\pm$ 0.12 & 0.25 $\pm$ 0.15\\ 
5.5 & 0.14 $\pm$ 0.02 & 0.16 $\pm$ 0.03 & 0.21 $\pm$ 0.01\\ 
6.5 & 0.19 $\pm$ 0.04 & 0.16 $\pm$ 0.05 & 0.17 $\pm$ 0.01\\ 
7.5 & 0.10 $\pm$ 0.06 & -0.19 $\pm$ 0.09 & 0.25 $\pm$ 0.02\\ 
8.5 & -0.13 $\pm$ 0.15 & -0.61 $\pm$ 0.21 & 0.26 $\pm$ 0.04\\ 
9.5 & 0.37 $\pm$ 0.36 & 0.10 $\pm$ 0.50 & 0.28 $\pm$ 0.10\\ 
\hline
&&&&UV&&&\\
\hline    
2.5 & 0.19 $\pm$ 0.01 & 0.41 $\pm$ 0.01 & 0.22 $\pm$ 0.01 && 0.61 $\pm$ 0.05 & -0.10 $\pm$ 0.01 & 0.15 $\pm$ 0.01\\ 
3.5 & 0.19 $\pm$ 0.01 & 0.35 $\pm$ 0.01 & 0.23 $\pm$ 0.01 && 0.94 $\pm$ 0.04 & -0.10 $\pm$ 0.03 & 0.20 $\pm$ 0.02\\
4.5 & 0.18 $\pm$ 0.01 & 0.20 $\pm$ 0.02 & 0.25 $\pm$ 0.01&& 0.31 $\pm$ 0.34 & -0.27 $\pm$ 0.10 & 0.07 $\pm$ 0.11\\ 
5.5 & 0.12 $\pm$ 0.03 & 0.06 $\pm$ 0.04 & 0.25 $\pm$ 0.01\\ 
6.5 & 0.25 $\pm$ 0.04 & 0.21 $\pm$ 0.05 & 0.19 $\pm$ 0.01\\ 
7.5 & 0.13 $\pm$ 0.06 & -0.17 $\pm$ 0.09 & 0.24 $\pm$ 0.02\\ 
8.5 & 0.17 $\pm$ 0.11 & -0.18 $\pm$ 0.15 & 0.22 $\pm$ 0.04\\ 
9.5 & 0.70 $\pm$ 0.20 & 0.50 $\pm$ 0.26 & 0.09 $\pm$ 0.08\\ 
\hline
	\label{table:size--mass}	
	\end{tabular}
\end{table*}

%%%%%%%%%%%%%%%%%%%%%%%%%%%%%%%%%%%%%%
\section{Size Distribution as a Function of Mass}\label{sec:size-dis}
%figure
\begin{figure*}
\centering
\includegraphics[width=2\columnwidth]{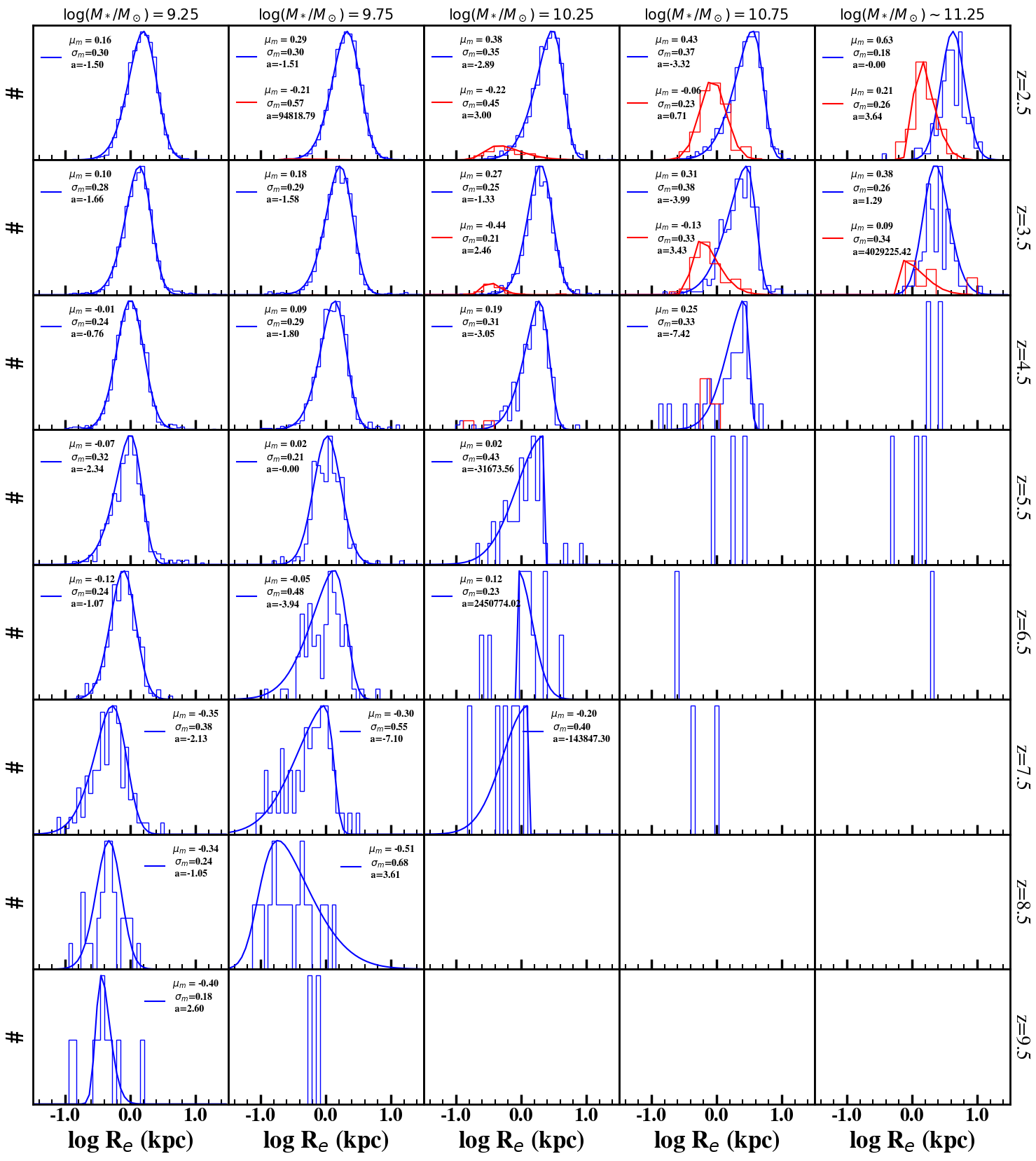}
\caption{Size distribution histograms for SFGs (blue) and QGs (red) in bins of stellar mass and redshift. The distributions are fit with a skewed normal distribution and the mean size $\mu_m$, scatter $\sigma_m$ and skewness parameter $a$ are labeled inside each panel. 
}
\label{img:size-hist}
\end{figure*} 

%%%%%%%%%%%%%%%%%%%%%%%%%%%%%%%%%%%%%%
\subsection{Size Distribution }\label{sec:size--massbins}
We present a complementary analysis of the evolution of the size–mass relation by examining size evolution as a function of stellar mass. Figure~\ref{img:size-hist} shows histograms of the size distributions categorized into various stellar mass and redshift bins. The five mass bins are $9<\log M_*/M_\odot<9.5$,  $9.5<\log M_*/M_\odot<10$,  $10<\log M_*/M_\odot<10.5$,  $10.5<\log M_*/M_\odot<11$, and  $11<\log M_*/M_\odot<11.5$. The size distributions are asymmetric. The distribution of SFGs skews toward the small side, and the skewness is more obvious for more massive galaxies and higher redshifts. Meanwhile, the distribution of QGs skews toward the large side. 
We characterize these size distributions by fitting a skew-normal distribution using the \texttt{Python} code \textit{scipy.stats.skewnorm}. The fitting results are shown as blue and red curves for SFGs and QGs, respectively, with the corresponding modeled mean size $\mu_m$, scatter $\sigma_m$, and skewness parameter $a$ displayed within each panel. The parameter $a$ determines the direction and magnitude of skewness, and the positive value suggests the right tail of the distribution is heavier and the peak of the distribution shifts leftward.  In some high-mass or high-redshift bins, fitting results are unavailable due to the limited sample sizes.

\begin{figure}
\includegraphics[width=\columnwidth, trim={0mm 0 0mm 0}, clip]{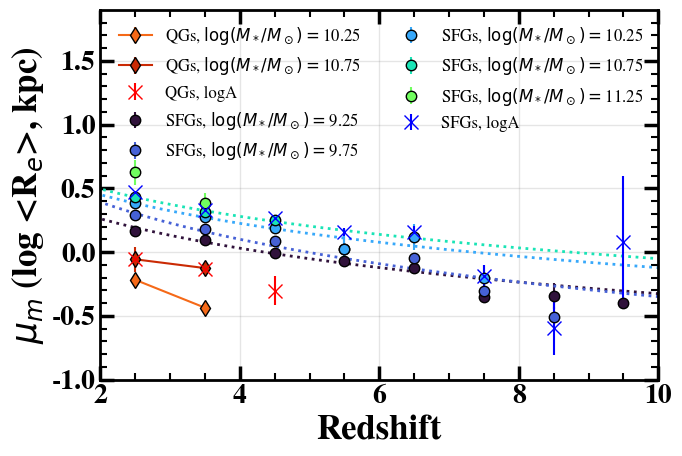}

\caption{Evolution of the modeled mean size, $\mu_m$, in different mass bins. Fitting results for SFGs in the form of $R_e\propto(1+z)^{\beta_m}$ are shown by dotted lines. The y-intercepts, $\log(A)$, inferred from the size--mass scaling relation are shown as crosses. }
\label{img:skwed-size-evolution}
\end{figure}

First, we show the evolution of the modeled mean size $\mu_m$ in different stellar mass bins in Figure~\ref{img:skwed-size-evolution}. For SFGs, we also fit the size evolution as $(1+z)^{\beta_m}$, and fitting results are shown as dotted lines. $\beta_m$ ranges from $-0.97$ to $-1.31$ and does not depend strongly on stellar mass. The size evolution in different mass bins is consistent with the size evolution inferred from the size--mass relation. For QGs, we did not fit the size evolution due to the limited sample size. Second, we find the scatter $\sigma_m$ is approximately 0.2-0.35 in most bins, but reaches $\gtrsim0.5$ at $z>6$ for mass bin $9.5<\log M_*/M_\odot<10$. Meanwhile, the skewness $a$ becomes more significantly negative compared to lower redshifts. The above characteristics suggest that our sample contains an additional larger fraction of small galaxies at high redshift compared to lower redshifts, see further discussion in the following section.

%figure

%%%%%%%%%%%%%%%%%%%%%%%%%%%%%%%%%%%%%%
\subsection{Tiny Blue Sources at $z>5$}\label{sec:tinyblue}
\begin{figure*}
\includegraphics[width=2\columnwidth, trim={80mm 0mm 30mm 0mm}, clip]{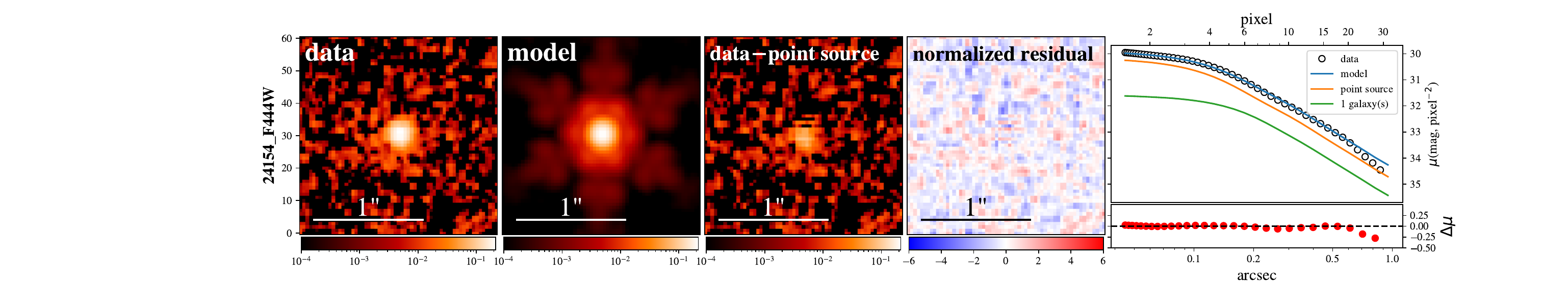}
\caption{Surface brightness profile fitting of an example unresolved source at $7<z<8$. This source is modeled by a S\'ersic profile plus a central point-like profile. }
\label{img:maybe-quasar}
\end{figure*}

For the lower redshift bins $z<5$, SFGs skew towards smaller size, which may be partly due to potential contamination from the QGs populations. This is because there is no strict bimodality between the two populations, whether in the color-color diagram or the sSFR criteria,  as also pointed out in \citet{vdW2014}.  However, this skewness still exists at $z>5$ where there are almost no significant effects from QGs. This excess of small galaxies may either reflect intrinsic differences in galaxy structure in the early Universe or possible contamination from AGN, despite our exclusion of X-ray AGN and LRDs. \citet{Morishita2024} also reported a number of blue compact sources in the size--mass plane, and those sources are suspected to be AGN candidates. We further examine the properties of these small galaxies, particularly those sources that are unresolved. We test the possible AGN nature of unresolved sources by performing a decomposition of of their surface brightness profiles, assuming a point source for the central AGN and a S\'ersic model for the extended host galaxy, following \citet{Ding2020}. We find that some unresolved sources exhibit morphological features indicative of AGN as presented in \citet{Ding2020}, such as source ID=24154 ($z=7.9$) as shown in Figure~\ref{img:maybe-quasar}. Further investigation is required to confirm the nature of these sources, which we will address in future work.

%figure

%%%%%%%%%%%%%%%%%%%%%%%%%%%%%%%%%%%%%%
\section{The Wavelength Dependence of Size and its Relation to Mass}\label{sec:size-wavelength}
Investigating the variation in galaxy size across different wavelengths offers valuable insights into the history of galaxy formation and the distribution of stellar populations. For example, observations at lower redshifts ($z<2$) indicate that galaxy sizes tend to appear smaller at longer wavelengths, suggesting an inside-out growth pattern where the inner regions of galaxies age first \citep{Suess2022}. At higher redshifts ($z>2$), due to the short cosmic time, a smaller stellar age difference is expected, hence smaller offsets between sizes at different wavelengths. 

At $z>3$, observations of galaxies' sizes in the rest-frame optical have only become feasible recently with the advent of JWST. Before this, many simulation studies made falsifiable predictions about the size evolution as a function of wavelength, mass, and luminosity \citep{Ma2018, Wu2020, Roper2022, Marshall2022, Costantin2023}.

%%%%%%%%%%%%%%%%%%%%%%%%%%%%%%%%%%%%%%
\subsection{$\Delta \log R_e$ as a Function of Stellar Mass}
\begin{figure*}
\includegraphics[width=2\columnwidth]{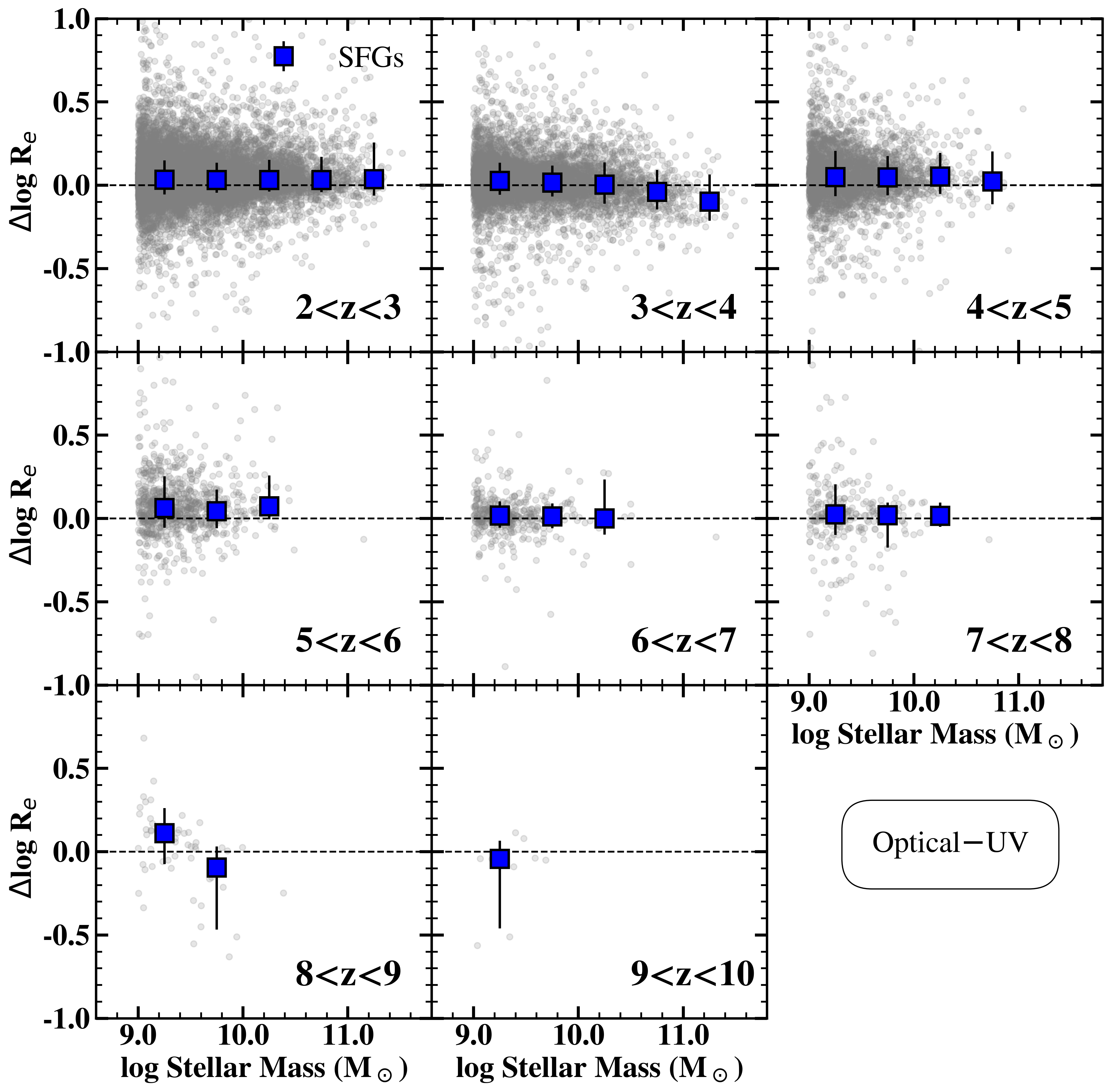}
\caption{Size difference, $\Delta \log R_{e}$, of SFGs between the rest frame optical and UV as a function of stellar mass in different redshift bins at $2<z<10$. The blue squares and error bars show the median value and its $1\sigma$ distribution. The dark dashed lines indicate the reference line $\Delta \log R_{e}=0$. }
\label{img:dsize}
\end{figure*}

In Figure \ref{img:dsize}, we show the size difference, $\Delta \log R_e$, between the rest-frame optical and UV for SFGs with secure measurements at both wavelengths as a function of stellar mass. We divide the galaxy sample into five mass bins as we did for studying size distribution, see Section~\ref{sec:size--massbins}. The squares represent the median value, while error bars represent the $1\sigma$ distribution and the dashed line represents $\Delta \log R_{e}=0$.

Generally, the optical sizes are comparable to the UV sizes. This is consistent with many JWST studies at $z>3$, which probe the similar or lower stellar mass range. Earlier work by \citet{Yang2022a} investigated the size as a function of wavelength at $z>7$ for galaxies with stellar mass $\sim \log(M_*/M_{\odot})=8\textendash9$ and found the size ratio to be almost unity. Similar results have also been reported by \citet{Morishita2024, Ono2024, Ormerod2024, Allen2024}. Simulations have also found similar results. For example, \cite{Costantin2023} measured the size for galaxies at $3<z<6$ in F200W and F365W from the \textsc{TNG50} cosmological simulation \citep{Nelson2019}, and found that the size ratio does not vary significantly at each mass bin.

In some cases, we find the optical sizes are slightly larger than UV toward the low mass end for some redshift bins, such as $4<z<5$ and $5<z<6$. One reason could be that UV light is heavily obscured by dust, such as observed in dusty star-forming galaxies \citep{Colina2023}. Another reason could be due to the fact that UV emission traces star-forming regions, which usually consist of smaller-scale clumpy structures as demonstrated by \textsc{FIRE} simulation \citep{Ma2018}.

Interestingly, we find that at some redshift bins, the more massive galaxies tend to have $\Delta logR_e<0$, meaning their optical sizes are smaller than their UV sizes. The median value of $\Delta logR_e$ is negative for massive galaxies with $log(M_*/M_{\odot})>10$ at $3<z<4$, as well as for the most massive galaxies at $8<z<9$. Our findings are qualitatively consistent with many simulations \citep{Marshall2022, Shen2024} at high-z. For example, \citet{Marshall2022} presented the observed images of high-z galaxies across different mass ranges and multiple bands (FUV, V, and H), using the  \textsc{BLUETIDES} cosmological hydro-dynamical simulation. They found that the intrinsic morphologies across different bands are similar and show no strong dependence on stellar mass. However, for more massive galaxies, the observed FUV images are more extended than their intrinsic counterparts. This is because dust is concentrated in more massive galaxies, causing the observed dust-attenuated FUV sizes to appear more extended. As a result, the observed FUV sizes are larger than the optical sizes.

The size difference, $\Delta \log R_e$, varies as a function of stellar mass, indicating that galaxies follow different growth tracks depending on their mass \citep{Shen2024}. In massive galaxies, the central regions are significantly affected by dust, which correlates with young stellar populations. This suggests that these galaxies undergo a compaction process, driven by rapid gas inflow toward the center, ultimately leading to a more compact structure and effective quenching. Further discussion of different evolutionary paths will be presented in Section~\ref{sec:lshape}.

%%%%%%%%%%%%%%%%%%%%%%%%%%%%%%%%%%%%%%
\subsection{Size--Mass Relation as a Function of Wavelength}
\begin{figure*}
\includegraphics[width=2\columnwidth]{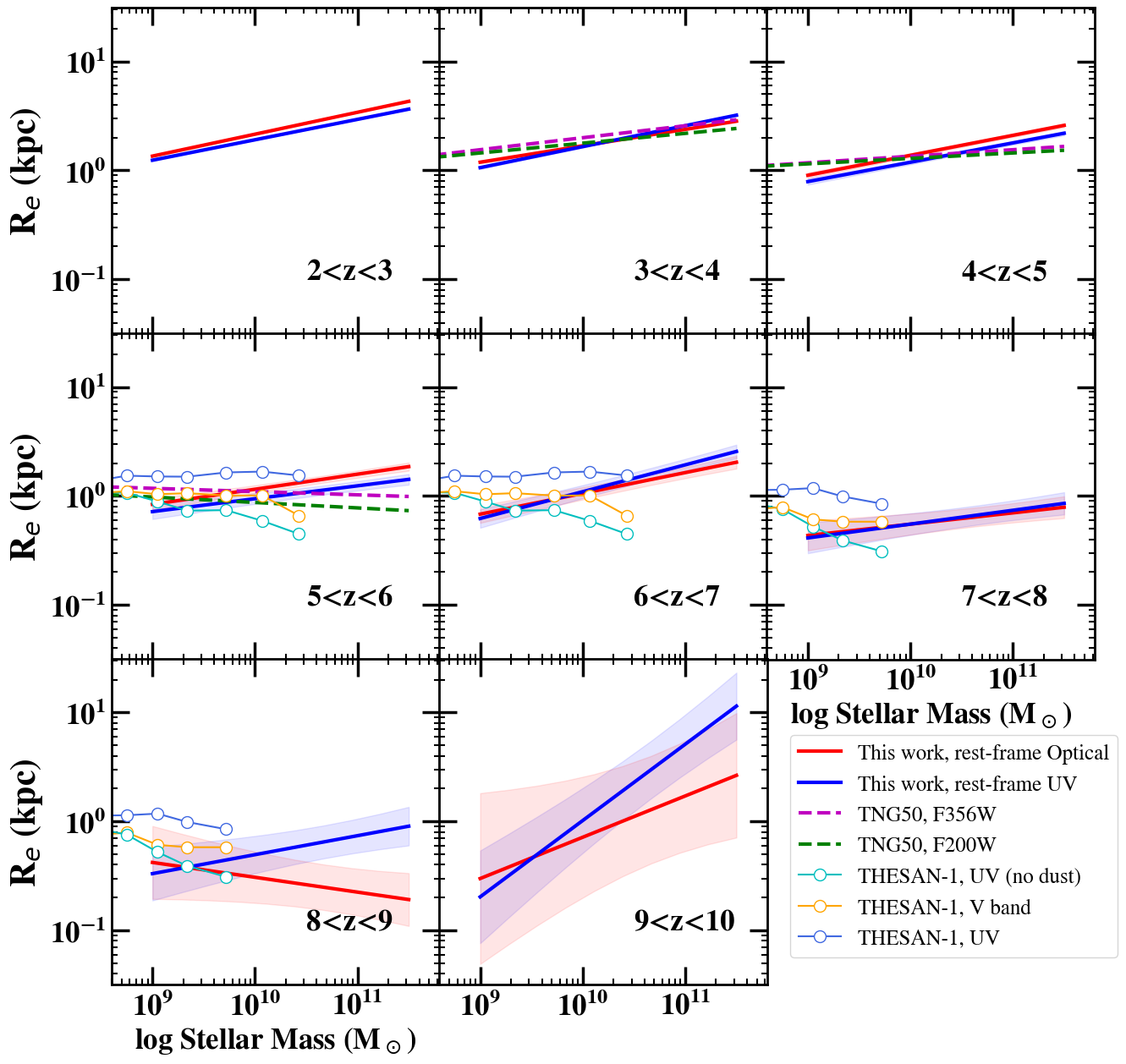}
\caption{
The size--mass relation for SFGs in several bins across a range of redshifts, $2<z<10$, comparing observations at rest-frame UV and optical wavelengths with results from the cosmological simulations TNG50 \citep{Costantin2023} and THESAN-1 \citep{Shen2024}.
}
\label{img:optical-vs-uv-sm}
\end{figure*}

We further investigate the variation of the size--mass relation as a function of wavelength. We adopt the same method used to derive the size--mass relation in the rest-frame UV as in optical, and the fitting results are presented in Table~\ref{table:size--mass}. In Figure~\ref{img:optical-vs-uv-sm}, we compare the best-fit scaling relation for SFGs derived from the UV (solid blue line) and optical (optical) at $2<z<10$. We find no significant differences between these two relations at $z<8$. This result is consistent with \cite{Costantin2023}, who predicted the size--mass relation at $z=3\textendash6$ using the TNG50 simulation dataset. Their predictions, shown as magenta (using the F356W filter for the optical) and green (using the F200W filter for the UV) dashed lines, align well with our observations. They also found that the observed slopes transition from positive to negative at $z\sim5$ for both wavelengths.

Despite their broad agreement, we find a discrepancy between the UV and optical, emerging at $z>8$, where the slope of the size--mass relation differs between rest-frame optical and UV wavelengths. Specifically, the slope is negative in the optical but positive in the UV.  The negative slope for the size--mass relation is counterintuitive. Interestingly, negative slopes in the size--mass relation have been reported from several simulations \citep{Roper2022, Marshall2022, Costantin2023, Shen2024}. For example, \cite{Roper2022}  analyzed the intrinsic and observed sizes of galaxies at $z>5$ using the \textsc{FLARES} simulation. They found that the intrinsic UV size-luminosity relation has a negative slope, conflicting with observational results \citep{Shibuya2015, Yang2022a}. This discrepancy is mitigated when dust attenuation is taken into account, as dust tends to be concentrated in galaxy cores, affecting observed sizes. After including the dust effect, the observed UV size will be much larger than its intrinsic value.

\citet{Roper2022} also present that the observed size-luminosity slopes decrease toward redder wavelengths including the effects of dust  (see their Figure 11). This trend is also supported by \cite{Marshall2022}. We observe a similar trend in our highest redshift bins, at $z>8$, though the sample size is limited in this range. A more straightforward comparison can be made with \cite{Shen2024}, who derived scaling relations in the UV (with and without dust effects) and optical V band at $z>5$. Their results demonstrate that dust attenuation increases observed UV sizes and has a more pronounced impact on massive galaxies, explaining the slope discrepancies.

%figure

%%%%%%%%%%%%%%%%%%%%%%%%%%%%%%%%%%%%%%
\section{Discussion}\label{sec:dis}
%%%%%%%%%%%%%%%%%%%%%%%%%%%%%%%%%%%%%%
\subsection{Size Ratio Between the Half-light Radius of Galaxy and Virial radius of Halo}
The angular momentum of the gas has long been considered one of the most important factors for regulating galaxy sizes. Both dark matter and diffuse gas acquire their angular momentum through the same process, tidal torques and merges. Similar to the stellar-to-halo mass relation (SHMR), there exists a stellar-to-halo radius relation, which describes the ratio between the half-mass or half-light radius of galaxies and the virial radius of the underlying dark matter halo \citep{Kravtsov2013}. The halo virial mass, $M_\text{vir}$, and radius, $R_\text{vir}$, are defined as that within the overdensity, $\Delta_\text{vir}$, times the critical density $\rho_\text{crit}$; therefore,  the virial radius is calculated by,
\begin{equation}
R_\text{vir} = \left(\frac{3M_\text{vir}}{4\Delta_\text{vir}\rho_\text{crit}}\right)^{1/3},
\end{equation}
where $\Delta_\text{vir} = 18\pi^2+82x-39x^2$ and $x=\Omega_m(z) -1$, and critical density, $\rho_\text{crit} = 3H(z)^2/8\pi G$. Following the disk formation model \citep{1998Mo},  the size of disk galaxies is proportional to the radius of the dark matter halo,
\begin{equation}
R_e/R_{vir} = \frac{1.68}{\sqrt{2}}\lambda f_j,
\label{eq:sizeratio}
\end{equation}
where $\lambda$ is the dark matter halo spin parameter and $f_j$ is the ratio of the specific angular momentum of the disk to that of the halo.

\begin{figure}
\includegraphics[width=\columnwidth]{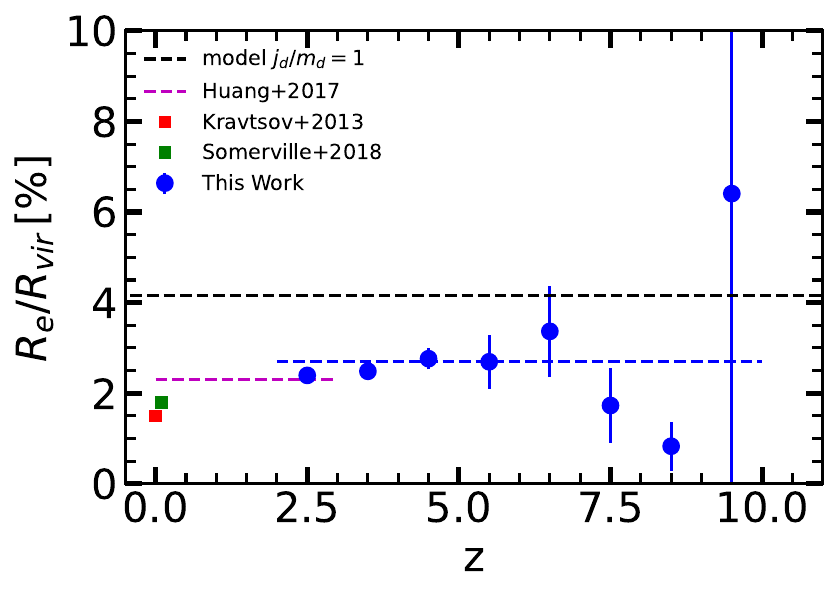}
\caption{Evolution of the ratio of the galaxy half-light radius to halo virial radius $R_e/R_\text{vir}$ for SFGs. The blue data points represent data in this work, with the blue dashed line indicating the median value of the size ratio, $R_e/R_\text{vir}=2.7\%$ at $z\sim2\textendash7$. The results obtained at lower redshifts are shown with a red square \citep{Kravtsov2013}, green square \citep{Somerville2018}, and magenta dashed line \citep{Kravtsov2013}, respectively.   The black dashed line shows the theoretical prediction assuming $f_j=1$, see Eq.~\ref{eq:sizeratio}.
 }
\label{img:stellar-size-to-halo-size}
\end{figure}

Following previous work, such as \citet{Shibuya2015}, for a galaxy with a certain stellar mass, we infer its underlying halo mass, $M_\text{vir}$, from the SHMR relation and subsequently calculate its virial radius. In this work, we adopt the SHMR relation from \citet{Shuntov2024} using the COSMOS-Web dataset, which performed abundance matching to estimate the relation across a large redshift range $0.2<z<12$ consistently, for the first time.  In Figure~\ref{img:stellar-size-to-halo-size}, we show the ratio between the half-light radius of the galaxy and virial radius of its host halo, $R_e/R_\text{vir}$, at $5\times10^{10}M_\odot$ as a function of redshift.  The $R_e/R_\text{vir}$ of SFGs is almost constant at $z\sim2\textendash7$ with a median value $\sim2.7$\%. The radius ratio decreases to $\sim1$\% at $7<z<8$ and $8<z<9$, and then increases to a larger value, albeit with large uncertainties.

We compare our results with those of other studies.  \citet{Kravtsov2013} found the radius ratio to be $1.5\%$ at $z\sim0$, and \citet{Somerville2018} found a similar result at $z=0.1$. \cite{Huang2017} reported the ratio is $2.3\%$ at $0<z<3$ and it remains unchanged using different SHMR functions, which is consistent with our results at $2<z<3$. \citet{Shibuya2015} also found a similar result at $z=0\textendash8$ but in the rest-frame UV, as well as \citet{Ono2025} which reported ratio $\sim0-4\%$ at $z>10$. Our finding shows that the radius ratio ranges approximately from $1\%$ to $3\%$.

The radius ratio can also be inferred from the spin parameter, as shown in Eq.~\ref{eq:sizeratio}. Based on simulations, the spin parameter is $\lambda\sim0.035$ and remains almost constant as a function of redshift, as demonstrated by \citet{RPA2016}. If we simply assume that the disk and halo have the same specific angular momentum, $f_j=1$, we obtain $R_e/R_\text{vir}\sim0.04$, as shown by the black dashed line in Figure~\ref{img:stellar-size-to-halo-size}. The observed radius ratio of SFGs is smaller than this theoretical expectation, indicating $f_j<1$. We also note that $R_e/R_\text{vir}$ tentatively decreases toward lower redshift since $z<7$. This result could suggest that, at early times, the angular momentum of galaxies is closer to that of their dark matter halos but decreases at lower redshifts. The growth of galaxies may even decouple from the growth of dark matter, as also speculated by \cite{Kravtsov2013}.

%figure

%%%%%%%%%%%%%%%%%%%%%%%%%%%%%%%%%%%%%%
\subsection{SFR Surface Density}
The SFR surface density, $\Sigma_\text{SFR}$, is defined as $\Sigma_\text{SFR} = 0.5 \times \text{SFR}/(\pi R_e^2)$. This parameter is directly linked to the surface density of gas, making $\Sigma_\text{SFR}$ a direct tracer of the local gas reservoir available for star formation \citep{Kennicutt1998}. Additionally, stellar feedback (e.g., from supernovae, stellar winds, and radiation) plays a critical role in regulating star formation by heating or expelling gas. Feedback efficiency depends on the local $\Sigma_\text{SFR}$. Studying the evolution of $\Sigma_\text{SFR}$ is crucial for understanding how efficiently galaxies convert gas into stars across cosmic time.

\subsubsection{$\Sigma_\text{SFR}$ Evolution}
\begin{figure}
\includegraphics[width=\columnwidth, trim={0mm 0 0mm 0}, clip]{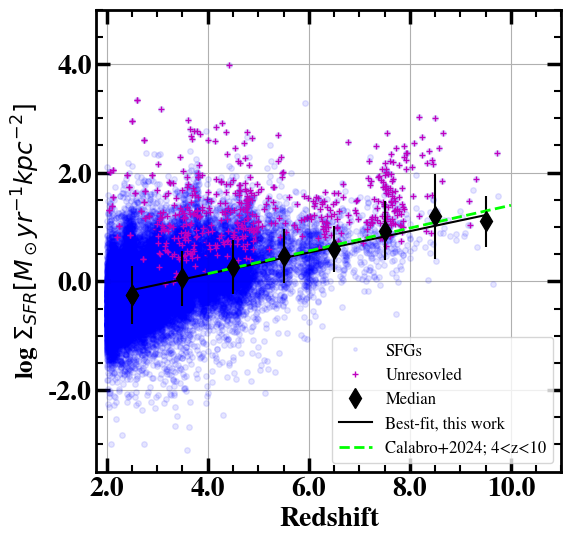}
\caption{
Evolution of the star formation rate surface density, $\Sigma_\text{SFR}$. Blue points represent individual star-forming galaxies (SFGs), while magenta crosses highlight unresolved sources. Black diamonds show the median $\Sigma_\text{SFR}$ in redshift bins, with error bars representing the $1\sigma$ error. The solid black line shows the best-fit linear relation, while the lime dashed line shows a comparison from \citet{Calabr2024} for $4<z<10$.
}
\label{img:sigmasfr}
\end{figure}

In Figure~\ref{img:sigmasfr}, we present the evolution of $\Sigma_\text{SFR}$ as a function of redshift at $2 < z < 10$.  All SFGs are shown as blue points, with their median values represented by black diamonds, while unresolved sources are highlighted by magenta crosses. We find that the median value of $\log\Sigma_\text{SFR}$ increases from $-0.5$ to 1.5 $M_\odot yr^{-1} kpc^{-2}$ at $2<z<10$. We fit a linear relation to $\log\Sigma_\text{SFR}$ and redshift, and the best-fit result is $\log\Sigma_\text{SFR} = (0.20\pm0.08)z+(-0.65\pm0.51)$ as shown by the black solid line. Our result agrees well with the best-fit results of the median $\log\Sigma_\text{SFR}$ reported by \citet{Calabr2024}, as shown by the lime dashed line. \citet{Calabr2024} include a sample with lower stellar mass reaching to $\log(M_*/M_{\odot})\sim8\textendash8.6$, and we will discuss the dependence of $\log\Sigma_\text{SFR}$ on stellar mass later in this section.

Furthermore, we observe a fraction of galaxies with a very high value of $\log \Sigma_\text{SFR} > 1.5$ in Figure~\ref{img:sigmasfr}.  Most of these extreme cases are unresolved sources, especially at $z>7$, this could be due to either those sources having efficient compact star-formation rates or hosting AGN as suggested by several works \citep{Morishita2024, Harikane2025, Ono2025}. Some of those compact sources have high-ionization emission lines \citep{Harikane2025}, which can be a strong hint of the existence of AGN, while our morphological decomposition also provides supportive evidence, see Section~\ref{sec:tinyblue}.

%figure

\subsubsection{$\Sigma_\text{SFR}\textendash M_*$ relation}
Because $\Sigma_\text{SFR}$ is closely tied to gas density, it has been proposed as a crucial parameter for understanding galaxy evolution, offering less ambiguity compared to the integrated star formation rate. For instance, \citet{Salim2023} demonstrated that the $\Sigma_\text{SFR}$ versus stellar mass $M_*$ plane at $0<z<2$, and they found the relation is flat locally, meaning that local main sequence galaxies exhibit similar $\Sigma_\text{SFR}$ values across a wide range of stellar masses, whereas the classical (s)SFR--$M_*$ relation is tilted. At higher redshift, $z \sim 2$, the relation slightly steepens toward the high-mass end but remains relatively flat.

\begin{figure}
\includegraphics[width=\columnwidth]{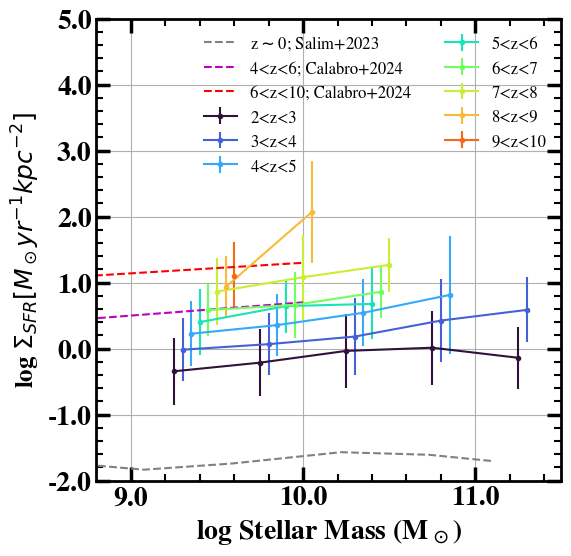}
\includegraphics[width=\columnwidth]{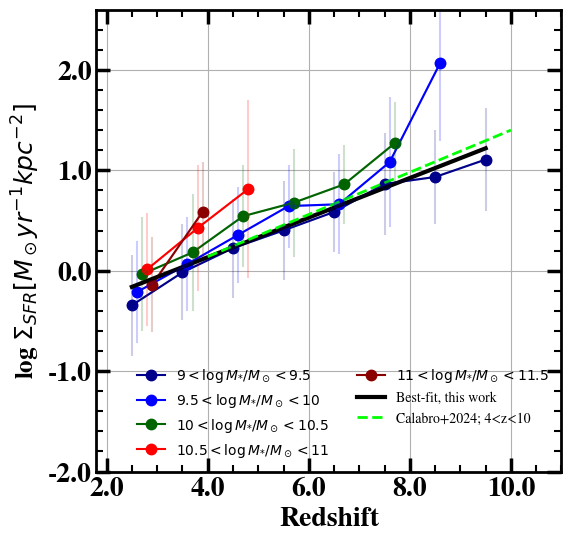}

\caption{ (Top) $\Sigma_\text{SFR}\textendash M_*$ relation across eight redshift bins at $z=2-10$. The gray dashed line shows the relation at $z\sim0$ \citep{Salim2023}, and magenta and red dashed lines represent the fitting results from \citet{Calabr2024} at $4<z<6$ and $6<z<10$, respectively. (Bottom) the evolution of $\Sigma_\text{SFR}$ in different redshift bins. The fitting results from this work (solid black line) and from \citet{Calabr2024} (lime dashed line) are as shown in Figure~\ref{img:sigmasfr}.
}
\label{img:sigmasfr_z_mass}
\end{figure}

In this work, we explore the $\Sigma_\text{SFR}\textendash M_*$ relation across eight redshift bins up to $z=10$ as shown in Figure~\ref{img:sigmasfr_z_mass} (top panel). We find a weak positive linear correlation between $\Sigma_\text{SFR}$ and $M_*$ at all redshift bins, with best-fit slopes of $0.12\pm0.31$, $0.31\pm0.31$, $0.36\pm0.56$, $0.28\pm0.73$, $0.28\pm0.56$, $0.40\pm0.64$, and $2.27\pm1.81$ for redshift bins between $z=2$ to $z=9$. While the slopes appear to increase toward higher redshifts, the large uncertainties make them consistent with zero within errors. At $9<z<10$, only one effective data point is available, preventing a reliable fit. These results are consistent with the flat $\Sigma_\text{SFR} \textendash M_*$ relation observed at $z\sim0$ (\citealt{Salim2023}, gray dashed line) and the mild positive trends reported by \citet{Calabr2024} at $4<z<6$ and $6<z<10$, as shown in magenta and red dashed lines, respectively. We further investigate the evolution of $\Sigma_\text{SFR}$ in different mass bins, shown in the bottom panel of  Figure~\ref{img:sigmasfr_z_mass}, comparing with the best-fit results of the median value of the entire sample and the result from \citet{Calabr2024}, as also shown in Figure~\ref{img:sigmasfr}. The evolution does not show a strong dependence on stellar mass.

%figure

%%%%%%%%%%%%%%%%%%%%%%%%%%%%%%%%%%%%%%
\subsection{Compaction and Quenching  }\label{sec:lshape}

%figure
\begin{figure*}
\includegraphics[width=1.9
\columnwidth]{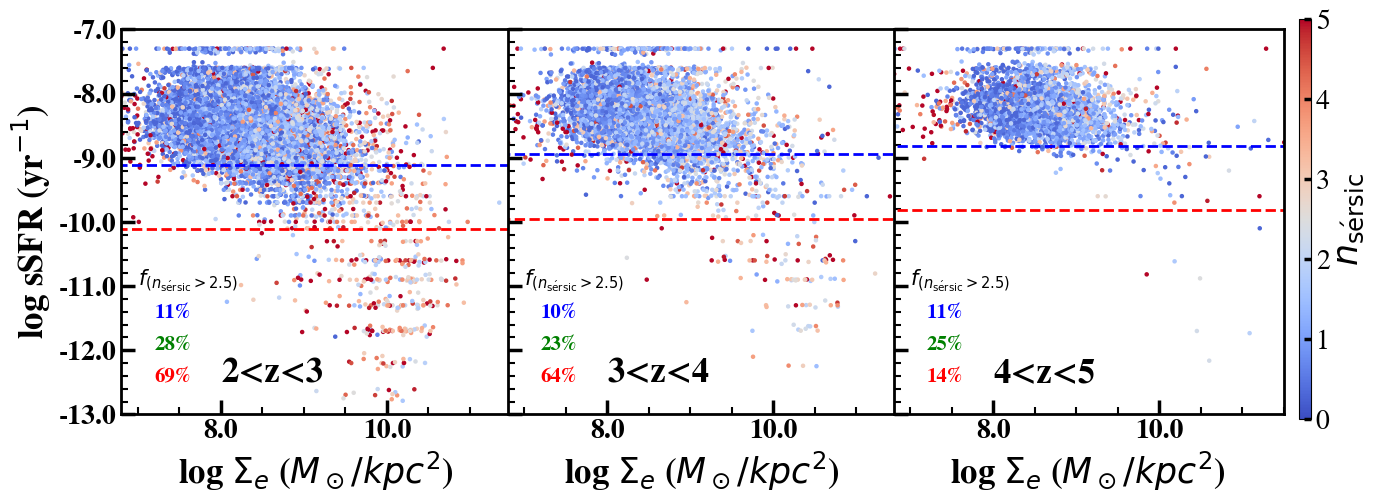}
\caption{The relation between stellar mass surface density ($\Sigma_{e}$) and specific star-formation rate ($\log_\text{sSFR}$), with points color-coded by S\'ersic index $n_{\text{s\'ersic}}$.  The red dashed line marks the threshold log(sSFR)$>0.2/t_{obs}$ separating star-forming and quiescent galaxies. The blue dashed line indicates log(sSFR)$>0.2/t_{obs}+1$, with galaxies between the two lines considered to be in a transition phase. Galaxies with $n_{\text{s\'ersic}}>2.5$ are considered bulge-dominated. The fractions of bulge-dominated galaxies in the star-forming, quiescent, and transition phases are labeled in blue, red, and green, respectively.}
\label{img:compaction}
\end{figure*} 

From the size--mass distribution shown in Figure~\ref{img:optical-sm}, at a given stellar mass, QGs are smaller leading to higher stellar mass surface densities than SFGs, suggesting a link between structural change and star formation quenching as part of galaxy evolution. \citet{Barro2017} found that SFGs and QGs follow distinct and tight correlations in stellar mass surface density, $\Sigma_e$, as a function of stellar mass. They suggested an evolutionary pathway where main-sequence SFGs grow inside-out, with both size and central density increasing over time while maintaining disk-like structures. Subsequently, galaxies undergo a compaction process that significantly enhances their central density, driving a transition toward bulge-like morphologies. Eventually, the star formation rate declines, leading to galaxy quenching. In this scenario, compaction is so tightly correlated with quenching that the galaxies will be quenched when reaching a $\Sigma_e$ threshold. 

In Figure~\ref{img:compaction}, we show the distribution of log(sSFR) and log$\Sigma_e$ to link star formation activity and mass distribution for SFGs and QGs at $2<z<5$. We show the threshold log(sSFR)$>0.2/t_{obs}$ with a red dashed line that separates star-forming and quiescent galaxies, and a blue dashed line for log(sSFR)$>0.2/t_{obs}+1$ indicating that the galaxies in the between are at a transition phase, such as green valley galaxies. The data points are colored by their S\'ersic index, $n_{\text{s\'ersic}}$. The $n_{\text{s\'ersic}}$ parameter describes the shape of a galaxy's surface brightness profile, where lower $n_{\text{s\'ersic}}$ values indicate disk-dominated galaxies and higher values indicate compact and bulge-dominated galaxies. 

We observe the characteristic L-shaped evolutionary track as described by \citet{Barro2017}. SFGs have lower $\Sigma_e$ and a disk-like morphology, i.e., the median value is $n_{\text{s\'ersic}}\sim1$. Galaxies then become compact with $\log\Sigma_e$ reaching $\sim9.5\textendash10M_{\odot}/kpc^2$, after that, galaxies become quenched. A similar threshold of compaction and quenching is also found at higher redshift $z>6$ as reported by simulations \citep{Shen2024}. We consider galaxies with $n_{\text{s\'ersic}}>2.5$ as bulge-dominated galaxies. To quantify the structural variation, we label the fraction of bulge-like galaxies at three phases, star-forming (blue), transitioning (green), and quenching (red) inside each panel.   

In all three redshift bins, the SFGs are dominated by disk-like morphologies and only a small fraction (10--11\%) has bulge-like structure. The fraction of bulge-like galaxies then increases to approximately 25\% at the transition phase. Those compact star-forming galaxies are likely to be quenching imminently, acting as progenitors of quiescent galaxies, which implies that the corresponding quenching mechanism is accompanied by the emergence of bulge-like structure in star-forming galaxies (see also \citet{Lang2014}).

Finally, quiescent galaxies are dominated by bulge-like features, consisting of $\sim70\%$ at $2<z<3$ and $3<z<4$. However, it is interesting that at $4<z<5$, only 14\% of quiescent galaxies have $n_{\text{s\'ersic}}>2.5$. The variation of bulge fraction in quiescent galaxies is compatible with the results of \citet{Huertas-Company2024}, where their galaxy morphologies are classified by a convolutional neural network. A similarly low value is also reported by \citet{Carnall2023}, who found a spectroscopically confirmed quiescent galaxy at $z=4.658$ with measured S\'ersic index $n_{\text{s\'ersic}}\sim2.3$. Additionally, the lower $n_{\text{s\'ersic}}$ value likely indicates that a fraction of quiescent galaxies have more disk-dominated structures, though other factors, such as an extended envelope surrounding the bulge, may also contribute. \cite{Ito2024} also investigated the S\'ersic index distribution of QGs at $z>3$, and they found that QGs in their sample have lower S\'ersic index $n_{\text{s\'ersic}}<2$. Our findings suggest that it is likely that the first generation of the QG population consists of a significant fraction of disk-like galaxies. 
      
The mechanisms driving compaction and quenching likely leave distinct morphological imprints. Simulations by \citet{Shen2024} also found the prevalence of disk-like morphologies at high redshifts as demonstrated in this work. They suggested that internal processes, such as disk instabilities, dominate over external triggers like mergers. They further show that gas inflow, driven by disk instability, enhances central density and triggers central star formation. This process is accompanied by a temporary increase in SMBH accretion and leads to eventual quenching, and the AGN host galaxies will remain disk-like. \citet{Onoue2024} analyzed two quasar host galaxies at $z>6$ and found that one of them exhibits little ongoing star formation while another is transitioning to a quiescent phase. However, determining AGN host galaxy morphology is still challenging, and we will further investigate this topic in future work.

\section{Summary and Conclusion}\label{sec:summary}

In this work, we analyze the evolution of galaxy sizes and their scaling relations with mass for SFGs and QGs across a wide redshift range ($2<z<10$).  The data used for this analysis come from COSMOS-Web, the largest area imaging survey conducted with JWST to date. We summarize our main findings as follows. 

\begin{itemize}
    
\item {\em Parameterized size--mass relation at the rest-frame optical up to $z=10$.} We parameterize the scaling relation between the half-light radius ($R_e$) and  stellar mass ($M_{*}$) of galaxies as $\log R_e/(kpc)= \log (A) + \alpha \log M_{*}/(5\times10^{10}M_{\odot})$. For SFGs, the slope, $\alpha$, at $2<z<8$ is approximately 0.20, showing no significant evolution with redshift.  At higher redshifts, the sample size is limited and we obtain the slopes $-0.13\pm0.15$ and $0.37\pm0.36$ at $8<z<9$ and $9<z<10$, respectively. The y-intercept $\log(A)$ decreases with redshift, following $R_e\propto(1+z)^\beta$, where $\beta=1.21\pm0.05$. The intrinsic scatter $\sigma_{\log Re}$ ranges between 0.2-0.3 dex for SFGs. For QGs, the slope $\alpha$, is steeper than that of SFGs, with values ranging from 0.5 to 0.8 at $2<z<5$. The y-intercept $\log(A)$ follows $R_e\propto(1+z)^{-0.81\pm0.26}$, and the intrinsic scatter ranges between 0.15 and 0.25 dex.

\item {\em Size and its relation with mass as a function of wavelength.} We have compared the size and its relation with mass in the rest-frame UV and optical. In summary, our findings indicate that the size--mass relation for SFGs is consistent between rest-frame UV and optical wavelengths at $z<8$. However, at $z>8$, the slope decreases from UV to optical, with a negative slope observed in the optical at $8<z<9$. Unlike predictions from cosmological simulations, we do not find an evident negative optical slope since $z>5$, which indicates the complex interplay between intrinsic galaxy properties and observational effects such as dust attenuation.

\item {\em Size ratio between galaxy and halo.} We find the size ratio, $R_e/R_{vir}$, between stellar light and its underlying dark matter halo using the latest SHMR result from COSMOS-Web. We find that $R_e/R_{vir}$ is almost consistent at $2<z<7$ with a median value 2.7\%, and slightly decreases at $z>7$. 

\item {\em SFR surface density as a function of redshift and stellar mass.} We find that $\Sigma_\text{SFR}$ increases monotonically with redshift following $\log\Sigma_\text{SFR} = (0.20\pm0.08)z+(-0.65\pm0.51)$, at $2<z<10$. We also investigate the $\Sigma_\text{SFR}\textendash M_*$ relation and find a weak positive correlation at all redshift bins. 

\item Galaxy quenching and morphological transformation. We also explored the link between galaxy structural changes and star formation quenching at $2<z<5$.  We find a threshold in stellar mass surface density $\log \Sigma_e=9.5\textendash10M_{\odot}/kpc^2$, beyond which galaxies become compact and eventually quench. Interestingly, quiescent galaxies at higher redshifts consist of a larger fraction sample with $n_{\text{s\'ersic}}<2.5$.

\end{itemize}

%\begin{acknowledgments}
This work is based on observations made with the NASA/ESA/CSA James Webb Space Telescope. The data were obtained from the Mikulski Archive for Space Telescopes at the Space Telescope Science Institute, which is operated by the Association of Universities for Research in Astronomy, Inc., under NASA contract NAS 5-03127 for JWST. These observations are associated with program \#1727. Support for this work was provided by NASA through grant JWST-GO-01727 awarded by the Space Telescope Science Institute, which is operated by the Association of Universities for Research in Astronomy, Inc., under NASA contract NAS 5-26555.

This work was made possible by utilizing the CANDIDE cluster at the Institut d'Astrophysique de Paris. The cluster was funded through grants from the PNCG, CNES, DIM-ACAV, the Euclid Consortium, and the Danish National Research Foundation Cosmic Dawn Center (DNRF140). It is maintained by Stephane Rouberol. Some of the measurements in this work are supported by World Premier International Research Center Initiative (WPI Initiative), MEXT, Japan.

%\end{acknowledgments}

\vspace{5mm}

\facilities{JWST(NIRCAM)}


\software{Galight \citep{Ding2020, Birrer2021},  
SourceXtractor++ \citep{Bertin2020, Kummel2020}, 
SAOImage DS9 \citep{ds9},
Numpy \citep{numpy}, 
Matplotlib \citep{matplotlib}, 
Scipy \citep{scipy}, 
Astropy\citep{astropy}
}




\bibliography{main}{}
\bibliographystyle{aasjournal}



\end{document}